\documentclass[journal,12pt,draftclsnofoot,onecolumn]{IEEEtran}
\IEEEoverridecommandlockouts
 \usepackage[utf8]{inputenc}
 \usepackage[T1]{fontenc}
 \usepackage{url}
 \usepackage{ifthen}
\usepackage{cite}
\usepackage{amsmath} 

\usepackage{amsthm}
\usepackage{amsfonts}
\usepackage{mathrsfs}
\usepackage{multirow}
\usepackage{subfigure}
\usepackage{bm}
\usepackage{algorithm}
\usepackage{graphicx}
\usepackage{epstopdf}
\usepackage{amssymb}
 \usepackage{diagbox}
 \usepackage{makecell}
 \usepackage{xhfill}

\allowdisplaybreaks[4]
\newcommand{\xfilll}[2][1ex]{%
	\dimen0=#2\advance\dimen0 by #1%
	\leaders\hrule height \dimen0 depth -#1\hfill%
}

\newenvironment{Proof}{
	{\indent\it Proof:}\;}{\hfill $\blacksquare$\par}
\newtheoremstyle{thry}
{3pt}
{3pt}
{}
{1em}
{}
{:}
{.5em}
{}
\theoremstyle{thry}

\newtheorem{theorem}{\emph{\textbf{Theorem}}}
\newtheorem{lemma}{\emph{\textbf{Lemma}}}
\newtheorem{corollary}{\emph{\textbf{Corollary}}}
\newtheorem{example}{{\textbf{Example}}}

\begin{document}
\title{Coding Theorems for Repetition and Superposition Codes over Binary-Input Output-Symmetric Channels \\
	\thanks{Corresponding author: Xiao Ma. This work was supported by the National Key R\&D Program of China~(Grant No.~2021YFA1000500). }
}

\author{
\IEEEauthorblockN{Yixin Wang and Xiao Ma, \textit{Member}, \textit{IEEE}}\\
\thanks{ Yixin Wang is with the School of Systems Science and Engineering, Sun Yat-sen University,  Guangzhou 510006, China~(e-mail:
	wangyx58@mail2.sysu.edu.cn)}
\thanks{  Xiao Ma is  with the Guangdong Key Laboratory of Information Security Technology, School of Computer Science and Engineering, Sun Yat-sen University,  Guangzhou 510006, China~(e-mail:
	maxiao@mail.sysu.edu.cn)}

}
\maketitle 

\begin{abstract}
This paper is concerned with a class of low density generator matrix codes~(LDGM), called  repetition and superposition~(RaS) codes, which have been proved to be capacity-achieving over
binary-input output-symmetric~(BIOS) channels in terms of bit-error rate~(BER). We prove with a recently proposed framework that the  RaS codes are also capacity-achieving over BIOS channels in terms of frame-error rate (FER). With this new framework, the theorem for the  RaS codes can be generalized to source coding and joint source and channel coding~(JSCC). In particular, we prove with this framework that the corresponding low-density parity-check~(LDPC) codes, as an enlarged ensemble of quasi-cyclic LDPC~(QC-LDPC) codes, can also achieve the capacity.  To further improve the iterative decoding performance, we consider the convolutional RaS~(Conv-RaS) code ensemble and prove it to be capacity-achieving over BIOS channels in terms of the first error event probability. The construction of Conv-RaS codes is flexible with rate~(defined as the ratio of the input length to the encoding output length) ranging from less than one ~(typically for channel codes) to greater than one~(typically for source codes), which can be implemented as a universal JSCC scheme, as confirmed by simulations. 
\end{abstract}
\begin{IEEEkeywords}
	Coding theorem,  low density generator matrix~(LDGM) codes, low density parity-check~(LDPC) codes, repetition and superposition~(RaS) codes 
	
\end{IEEEkeywords}

\section{Introduction}
\par  Low density parity-check (LDPC) codes, which were invented by Gallager in the early 1960s~\cite{1962GallagerLDPC}, are a class of linear codes with sparse parity-check matrices.  It has been proved~(numerically by density evolution~(DE)) in~\cite{Richardson2001DE,Richardson2001LDPC} that LDPC codes can be a class of capacity-approaching under iterative message passing decoding over a broad class of channels.  According to the construction approaches, LDPC codes can be roughly classified into two categories: pseudorandom~(or random-like) codes~\cite{MacKay1999sparse,Richardson2001LDPC,hu2005regular} and structured codes~\cite{kou2001low,fossorier2004quasicyclic,song2009QC,diao2016QC}. As a special class of structured LDPC codes, quasi-cyclic LDPC (QC-LDPC) codes are generally specified by an array of circulants, say, circulant permutation matrices (CPMs)~\cite{fossorier2004quasicyclic}.
QC-LDPC codes are known for their efficient encoder and decoder hardware implementation, fast decoding convergence, and low error floors. Due to these advantages, they have been extensively studied and widely applied~\cite{IEEE2021WIFI,CCDS2017,5GeMBB}. However, to the best of our knowledge, the asymptotic performance of QC-LDPC codes is mainly analyzed from the extrinsic information transfer~(EXIT) charts~\cite{2001EXIT,2007LivaPEXIT}, and no proof is available in the literature that the QC-LDPC code ensembles are capacity-achieving.

\par   As duals of LDPC codes, LDGM are a class of linear codes with sparse generator matrices.  The LDGM ensembles can be constructed by degree distributions~\cite{sourlas1989spin,cheng1996some,Luby2002LT,Montanari2005Tight}, or based on Bernoulli process~\cite{Kakhaki2012LDGM,Ma2016Coding,Cai2020SCLDGM,Wang2022}. In~\cite{Kakhaki2012LDGM}, the non-systematic LDGM code ensemble with generator matrix defined by the Bernoulli process was proved to be capacity-achieving over binary symmetric channels~(BSCs).  The systematic LDGM codes with generator matrices defined by Bernoulli process, called Bernoulli generator matrix (BGM) codes, were proved to be capacity-achieving over binary-input output-symmetric~(BIOS) memoryless channels in terms of bit-error rate~(BER)~\cite{Ma2016Coding,Cai2020SCLDGM} and in terms of frame-error rate~(FER)~\cite{Wang2022}. The performance of LDGM codes is not that good as channel codes due to their high error floors caused by the low weight codewords~\cite{cheng1996some}, which can also be observed in systematic BGM code ensemble because of the randomness of generation.  However, the error floors can be lowered down by concatenating with high-rate outer codes~\cite{2003Approaching,Lopez2007Serially}. It has been shown that the raptor codes~\cite{Shokrollahi2006Raptor}~(concatenation of outer linear block codes and inner LT codes) can achieve the capacity of the binary erasure channels~(BECs).

\par  To improve the performance, the convolutional version of LDPC codes have been investigated and studied~\cite{jimenez1999SCLDPC,kudekar2011threshold,kudekar2013spatially,kumar2014threshold} in the past decades, which are also referred to as spatially coupled LDPC (SC-LDPC) codes. The
spatially coupled codes exhibit a threshold saturation phenomenon~\cite{kudekar2011threshold}, which has been proved for BECs~\cite{kudekar2011threshold} and generalized to BIOS memoryless channels~\cite{kudekar2013spatially}. Thus, the SC-LDPC codes can achieve capacity universally over BIOS memoryless channels~\cite{kudekar2013spatially,kumar2014threshold}.  The spatial coupling technique can also be applied to the LDGM codes. In~\cite{kumar2014threshold}, spatially coupled LDGM codes were also proved to achieve the capacity of BIOS channels. 

\par In this paper, we use the code enlargement technique~\cite{Gallager1968,1999ShulmanRandom} and the framework proposed in~\cite{Wang2022} to prove the coding theorem for repetition and superposition~(RaS) code ensemble, which takes the LDGM code ensemble with an array of CPMs as members. This theorem can be generalized to source coding and joint source~(JSCC). We can also prove that the corresponding parity-check code ensemble can achieve the capacity. To further improve the iterative decoding performance, we consider the convolutional RaS~(Conv-RaS) code ensemble and prove the coding theorem. The main contributions of this paper are summarized
as follows.
\begin{itemize}
	\item [1)] \textbf{Coding Theorem for RaS Code Ensemble:} We prove that the LDGM code ensemble, called RaS code ensemble, can be capacity-achieving over BIOS channels in terms of FER. The coding theorem can be generalized to source coding and JSCC. We also prove that  the corresponding LDPC code ensemble, as an enlarged ensemble of QC-LDPC codes, are capacity-achieving.
	\item [2)] \textbf{Coding Theorem for Conv-RaS Code Ensemble:} To further improve the iterative decoding performance, we introduce the Conv-RaS code ensemble and give the coding theorem. Our proof suggests that convolutional LDPC codes can achieve the capacity. 
	\item [3)] \textbf{Flexibility and Universality of Conv-RaS codes:} We treat the Conv-RaS coding as an asynchronous multiple access channel system with a tunable delay. As a result, the construction of Conv-RaS codes is flexible with rate~(defined as the ratio of the input length to the encoding output length) ranging from less than one ~(typically for channel codes) to greater than one~(typically for source codes), which can be implemented as a universal JSCC scheme, as confirmed by simulations. 
\end{itemize}
\par The rest of the paper is organized as follows. In Section~\ref{sec2}, we introduce the code enlargement and describe the framework. In Section~\ref{sec3}, we give the main results of this paper, including the coding theorems for the RaS code ensemble and the Conv-RaS code ensemble and the flexibility and universality of the Conv-RaS codes. In Section~\ref{sec4}, we provide the lemmas and theorems to prove the main results and give the detailed proofs of the coding theorems. In Section~\ref{sec5}, we provide the simulation results to show the flexibility and universality of the Conv-RaS codes. Section~\ref{sec6} concludes this paper.

\par In this paper, a random variable is denoted by an upper-case letter, say $X$, whose realization is denoted by the corresponding lowercase letter $x\in \mathcal{X}$. We use $P_{X}(x)$, $x\in \mathcal{X}$ to represent the probability mass~(or density) function of a random discrete~(or continuous) variable. For a vector of length $m$, we represent it as $\bm x=(x_0,x_1,\cdots,x_{m-1})$. We also use $\bm x^m$ to emphasize the length of $\bm x$. We denote by $\mathbb{F}_2=\{0,1\}$ the binary field. We denote by $\log$  the base-2 logarithm and by $\exp$  the base-2 exponent. We denote by $[0:k]$ the set $\{0,1,\cdots,k\}$.
\section{Motivation and System Model}\label{sec2}
\subsection{Motivation}
The parity check matrix of the QC-LDPC code ensembles with arrays of CPMs are of the form~\cite{fossorier2004quasicyclic}
\begin{equation*}
	\mathbf{H} = \left[
	\begin{array}{cccccccc}
		\mathbf{H}_{0,0} & \mathbf{H}_{0,1} &  \cdots    & \mathbf{H}_{0,n-1}\\
		\mathbf{H}_{1,0}	& \mathbf{H}_{1,1} & \cdots  & \mathbf{H}_{1,n-1}\\
		\vdots	&  \vdots       & \vdots  & \vdots   \\
		\mathbf{H}_{m-1,0}	& \mathbf{H}_{m-1,1}        & \cdots & \mathbf{H}_{m-1,n-1}    \\
	\end{array}
	\right]\text{,}
\end{equation*}
where $\mathbf{H}_{i,j}$ is a $B\times B$ circulant permutation matrix or zero matrix. For the non-zero matrix $\mathbf{H}_{i,j}$ in $\mathbf{H}$, the columns have certain constraints, say, being distinct. Due to the constraints, it is difficult to prove the coding theorem for the code ensembles even though the CPMs or zero matrices in $\mathbf{H}$ are chosen randomly. However, if  the parity check matrices of the QC-LDPC code ensembles are with sufficiently large $m$ and $n$, we can then apply a random interleaver of size~$nB$ to permute the columns of $\mathbf{H}$, resulting in the following parity-check matrix $\mathbf{H}^\prime$ with 
\begin{equation*}
	\mathbf{H}^\prime = \left[
	\begin{array}{cccccccc}
		\mathbf{H}_{0,0}^\prime & \mathbf{H}_{0,1}^\prime &  \cdots    & \mathbf{H}_{0,n-1}^\prime\\
		\mathbf{H}_{1,0}^\prime	& \mathbf{H}_{1,1}^\prime & \cdots  & \mathbf{H}_{1,n-1}^\prime\\
		\vdots	&  \vdots       & \vdots  & \vdots   \\
		\mathbf{H}_{m-1,0}^\prime	& \mathbf{H}_{m-1,1}^\prime        & \cdots & \mathbf{H}_{m-1,n-1}^\prime    \\
	\end{array}
	\right]\text{.}
\end{equation*}
In this case, the columns of the sub-matrix $\mathbf{H}_{i,j}^\prime$ can be treated approximately as independently distributed. Motivated by this observation, we may assume that $\mathbf{H}_{i,j}^\prime$ is a matrix of size $B\times B$ with each column drawn independently and uniformly from $\mathcal{B}=\{\bm v^B \in \mathbb{F}_2^B | W_H(\bm v^B )\leq 1\}$, the collection of all binary column vectors of weight $0$ or $1$, where $W_H(\cdot)$ denotes the Hamming weight function. The new code ensemble takes as members the QC-LDPC codes. In this paper, we use the above method to enlarge the code ensemble and prove the enlarged ensemble is capacity-achieving in channel coding, source coding and JSCC with the following framework.
\subsection{System Model}
\begin{figure}[t]
	\centering
	\includegraphics[width=0.5\textwidth]{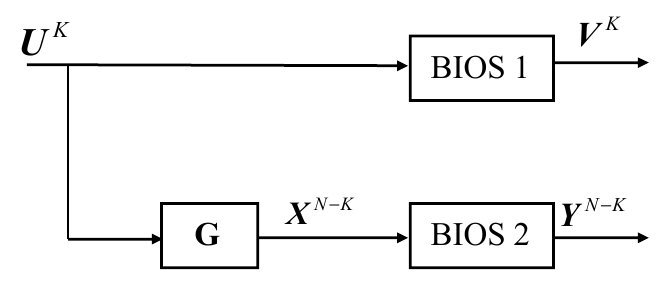}
	\caption{A  system model with systematic linear coding. }
	\label{systemm model}
\end{figure}
\par  We consider a system model that is depicted in Fig.~\ref{systemm model}~\cite{Wang2022}, where~$\bm U^{K}\in \mathbb{F}_2^{K}$ is referred to as the message bits to be transmitted, ${\mathbf{G}}$ is a  binary matrix of size $K\times (N-K)$ and~$\bm X^{N-K}=\bm U^K{\mathbf{G}}\in \mathbb{F}_{2}^{N-K}$ is referred to as parity-check bits corresponding to $\bm U^K$. The message bits $\bm U^K$ and the parity-check bits $\bm X^{N-K}$ are transmitted through two~(possibly different) BIOS channels, resulting in~$\bm V^K$ and~$\bm Y^{N-K}$, respectively. A  BIOS  memoryless channel is characterized by an input~$x \in \mathcal{X}=\mathbb{F}_{2}$, an output set~$\mathcal{Y}$~(discrete or continuous), and a conditional probability mass (or density) function\footnote{If the context is clear, we may omit the subscript of the probability mass (or density) function.}$\{P_{Y|X}(y|x)\big| x\in\mathbb{F}_2, y\in\mathcal{Y}\}$ which satisfies the symmetric condition that $P_{Y|X}(y|1)=P_{Y|X}(\pi(y)|0)$ for some mapping~$\pi~: \mathcal{Y}\rightarrow\mathcal{Y}$ with $\pi^{-1}(\pi(y))=y$. For simplicity, we assume that the BIOS channels are memoryless, meaning that~$P_{\bm{V}|\bm{U}}(\bm u|\bm v)=\prod\limits_{i=0}^{K-1}P_{V|U}(v_i|u_i)$ and~$P_{\bm{Y}|\bm{X}}(\bm y|\bm x)=\prod\limits_{i=0}^{N-K-1}P_{Y|X}(y_i|x_i)$. For a Bernoulli input~$X$, we can define the mutual information~$I(X; Y)$.  The channel capacity of the BIOS channel is given by $C = I(X; Y)$ with $X$ being a uniform binary random variable with $P_X(0) = P_X(1) = 1/2$.
\par  The task of the receiver is to recover $\bm U^K$  from $(\bm V^K,\bm Y^{N-K})$. The BER   is defined as~$\textbf{E}[W_H(\hat{\bm U}^K+\bm U^K)]/K$, where~$\textbf{E}[\cdot]$ denotes the expectation of the random variable and~$\hat{\bm U}^K$ is the estimate of~$\bm U^K$ from decoding.  The FER is defined as ${\rm Pr}\{\hat{\bm U^K}\neq \bm U^K\}$. 
\subsection{Repetition and Superposition~(RaS) code ensemble}
\par To prove the coding theorem for the enlarged QC-LDPC code ensemble, we introduce the following code ensemble.
\par \textbf{Repetition and Superposition~(RaS) code ensemble~\cite{Ma2016Coding}}:~The generator matrix has the form~$[\mathbf{I}~\mathbf{G}]$ of size $K\times N$ with $K=k(m+1)$ and $N=n(m+1)$, where

\begin{equation*}
	\mathbf{G} = \left[
	\begin{array}{cccccccc}
		\mathbf{G}_{0,0} & \mathbf{G}_{0,1} &  \cdots    & \mathbf{G}_{0,m}\\
		\mathbf{G}_{1,0}	& \mathbf{G}_{1,1} & \cdots  & \mathbf{G}_{1,m}\\
		\vdots	&  \vdots       & \vdots  & \vdots   \\
		\mathbf{G}_{m,0}	& \mathbf{G}_{m,1}        & \cdots & \mathbf{G}_{m,m}    \\
	\end{array}
	\right]\text{,}
\end{equation*} 
and $\mathbf{G}_{i,j}$ is a random matrix of size $k\times (n-k)$ with each column drawn independently and uniformly from $\mathcal{K}=\{\bm v^k \in \mathbb{F}_2^k | W_H(\bm v^k )\leq 1\}$, the collection of all binary column vectors of weight $0$ or $1$. A message vector of length $K$ can be written as $\bm u^{K}=(\bm u^{(0)},\bm u^{(1)},\cdots,\bm u^{(m)})$  where $\bm u^{(i)}\in \mathbb{F}_2^k~(0\leq i\leq m)$ is referred to as a sub-block of $\bm u^{K}$. The weight profile of a message sequence $\bm u^K$ is defined as $\bm w^{m+1}=(w_0,w_1,\cdots,w_m)$ with $w_i=W_H(\bm u^{(i)})\in [0:k]~(0\leq i\leq m)$.

\par
\begin{figure}[t]
	\centering
	\includegraphics[width=0.5\textwidth]{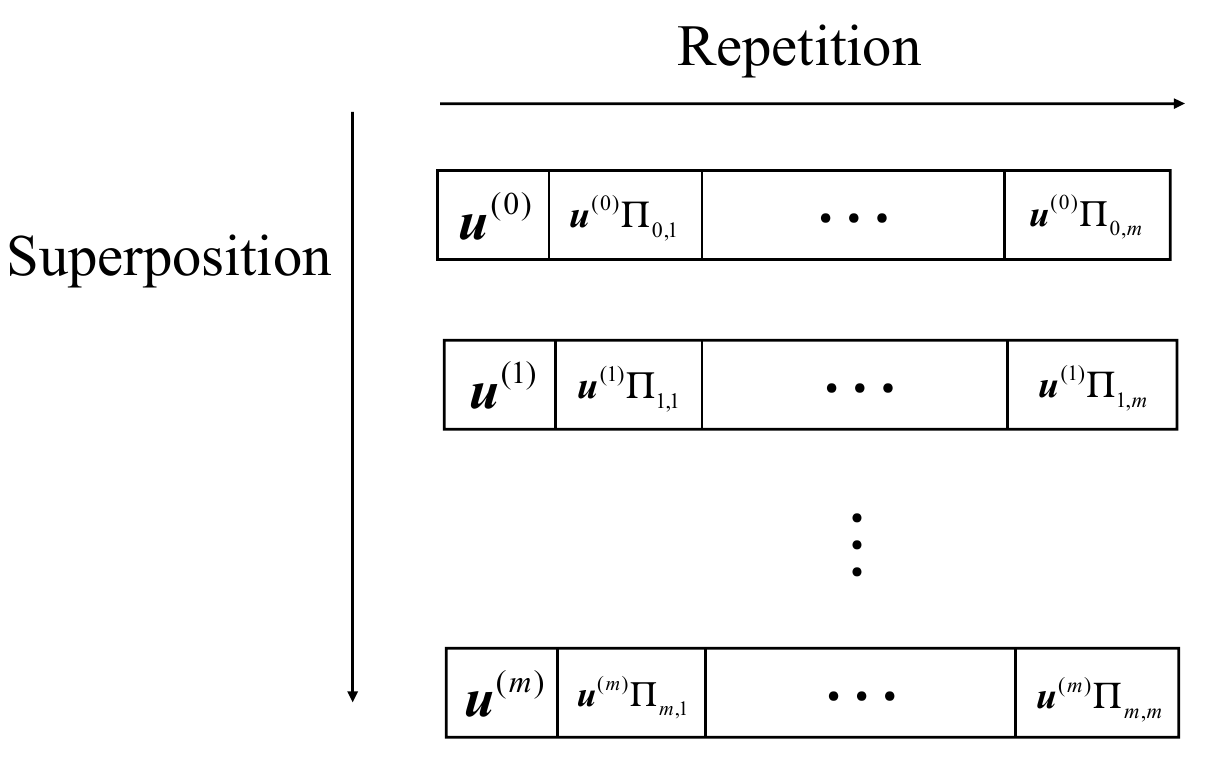}
	\caption{The encoding of the RaS codes, where $\Pi_{i,j}(0\leq i\leq m, 1\leq j\leq m)$ is a random permutation matrix. }
	\label{RaS}
\end{figure}

 \textbf{Remark:} In the case when the matrix $\mathbf{G}$ is an~(interleaved) array of CPMs,  the encoding is equivalent to repeating and interleaving $\bm u^{(i)}$ multiple times and then performing superposition, as illustrated in Fig.~\ref{RaS}. This is why we call the code as \textbf{Repetition and Superposition code}. 

\section{Main Results}\label{sec3}

\subsection{Coding Theorem for Systematic RaS Code Ensemble}
\label{block_ensemble}

\begin{theorem}
	\label{theorem_for_block_code}
	Consider the RaS code ensemble defined by the generator matrices of the form~$[\mathbf{I}~\mathbf{G}]$. Let $k$ and $n$ be two positive integers  such that $k/(n-k)<I(X;Y)/H(U|V)$. For an arbitrarily small positive number $\epsilon$, one can always find a sufficiently large integer~$m_0$ such that, for all $m \geq m_0$, there exists a matrix $\mathbf{G}$ of size $(m+1)k\times(m+1)(n-k)$ and a decoding algorithm satisfying the error rate ${\rm FER} \leq \epsilon$.
\end{theorem}
From Theorem~\ref{theorem_for_block_code}, we have the following corollaries.
\begin{corollary}
	As source coding, the RaS code ensemble can achieve the entropy $H(U)$.
\end{corollary}
\begin{Proof}
	For source coding, BIOS~$1$ is a totally erased channel~\footnote{By the totally erased channel we mean an erasure channel with erased probability~$1$.} and BIOS~$2$ is noiseless. In this scenario, the condition  $k/(n-k)<I(X; Y)/H(U|V)$  in Theorem~\ref{theorem_for_block_code} is equivalent to the conventional condition $(n-k)/k>H(U)$, where $(n-k)/k$ is the code rate for the source coding. This can be verified by noting that $H(U|V)=H(U)$ and $I(X; Y)=1$.
\end{Proof}
\begin{corollary}
	As channel coding, the RaS code ensemble can achieve the capacity $C$.
\end{corollary}
\begin{Proof}
	For channel coding,  BIOS~$1$ and BIOS~$2$ are the same. In this scenario, the condition $k/(n-k)<C/H(U|V)$ is equivalent to the conventional condition $k/n<C$, where $k/n$ is the code rate for channel coding. This can be verified by noting that $C=1-H(X|Y)$ and~$H(U|V)=H(X|Y)$.
\end{Proof}

\begin{corollary}
	\label{corollary_JSCC}
	For JSCC, the RaS code ensemble can achieve the limit of the minimum transmission ratio~(LMTR) $H(U)/C$~\cite{Csizar2011}.
\end{corollary}
\begin{Proof}
	For JSCC, BIOS 1 is a totally erased channel, and BIOS~$2$ has noise. In this scenario, the above theorem states that the transmission ratio  $(n-k)/k$ can be arbitrarily close to  the LMTR $H(U)/C$. 
\end{Proof}
\par A parity-check code ensemble associated with the RaS code ensemble is defined by  $\mathscr{C}=\{\bm u\in \mathbb{F}_2^{K}|\bm u\mathbf{G}=\bm 0\}$, where $\mathbf{G}$ is the matrix defined in Sec.~\ref{block_ensemble} of size $K\times (N-K)$. It can be seen that the parity-check code ensemble takes the QC-LDPC codes as members. So the following corollary can be viewed as a coding theorem for the enlarged QC-LDPC code ensemble.
\begin{corollary}
	The parity-check code ensemble associated with the RaS code ensemble can achieve the capacity $C$ of the BIOS channel for Bernoulli source $P(0)=P(1)=1/2$.
\end{corollary}
\begin{Proof}
	According to the definition, the parity-check matrix of the parity-check code ensemble is then given by~$\mathbf{H}=\mathbf{G}^T$.
	We only need to find ways to tell the decoder the noiseless parity-check vector $\bm y^{N-K} = \bm x^{N-K}$. This can be readily resolved by imposing a constraint that only those $\bm u$ with $\bm u\mathbf{G}=\bm 0$ can be legally transmitted. In this case, the BIOS~$1$ is a (possibly) noisy channel and the BIOS~$2$ is a noiseless channel. Hence, $I (X;Y) / H(U|V) = 1/ H(U|V)$ and $(n-k) / k$ can be arbitrarily close to $H(U|V)$. Equivalently, the code rate  $R = (k-{\rm rank}(\mathbf{G}))/k \geq  (2k-n) / k$  can be arbitrarily close to $1-H(U|V)=C$, where $C$ is the capacity of BIOS~1.
\end{Proof}
\subsection{Coding Theorem for Convolutional RaS~(Conv-RaS) Code Ensemble}
We consider the linear coding framework in Fig.~\ref{systemm model} with BIOS~$1$ totally erased and consider the convolutional RaS code ensemble, which have better performance under iterative decoding than their block counterpart.

\par \textbf{Convolutional Repetition and Superposition~(Conv-RaS) code ensemble~\cite{Ma2016Coding}}:~The generator matrix $\mathbf{G}$ has the form   
\begin{equation}
	\label{SC_matrix}
	\mathbf{G} = \left[
	\begin{array}{cccccccc}
		\mathbf{G}_{0,0} & \mathbf{G}_{0,1} &  \cdots    & \mathbf{G}_{0,m}   &     & \\
		& \mathbf{G}_{1,1} & \mathbf{G}_{1,2} &  \cdots  & \mathbf{G}_{1,m+1} &\\
		&         & \ddots  & \ddots    & \ddots  & \ddots \\
	\end{array}
	\right]\text{,}
\end{equation}
where $\mathbf{G}_{i,j}$ is a random matrix of size $k\times (n-k)$ with each column drawn independently and uniformly from $\mathcal{K}=\{\bm v^k \in \mathbb{F}_2^k | W_H(\bm v^k )\leq 1\}$, the collection of all binary column vectors of weight $0$ or $1$.

\par \emph{First error event probability:}  Define the error event  $\{\bm u^{(0)}\neq \bm \hat{\bm u}^{(0)}\}$ of the first sub-block $\bm u^{(0)}$ as first error event and the first error event probability is  ${\rm Pr}\{\bm u^{(0)}\neq \bm \hat{\bm u}^{(0)}\}$.

\begin{theorem}
	\label{theorem_for_conv}
	Consider Conv-RaS code ensemble defined above. Let $k$ and $n$ be two positive integers such that $k/(n-k)<I(X;Y)/H(U)$.  For an arbitrarily small positive number $\epsilon$, one can always find a sufficiently large integer $m_0$ such that, for all  $m>m_0$, there exists a matrix $\mathbf{G}$ in~\eqref{SC_matrix} and a decoding algorithm satisfying the  first error event probability ${\rm Pr}\{\bm u^{(0)}\neq \hat{\bm u}^{(0)}\}\leq\epsilon$.
\end{theorem}

\begin{figure}[t]
	\centering
	\includegraphics[width=0.68\textwidth]{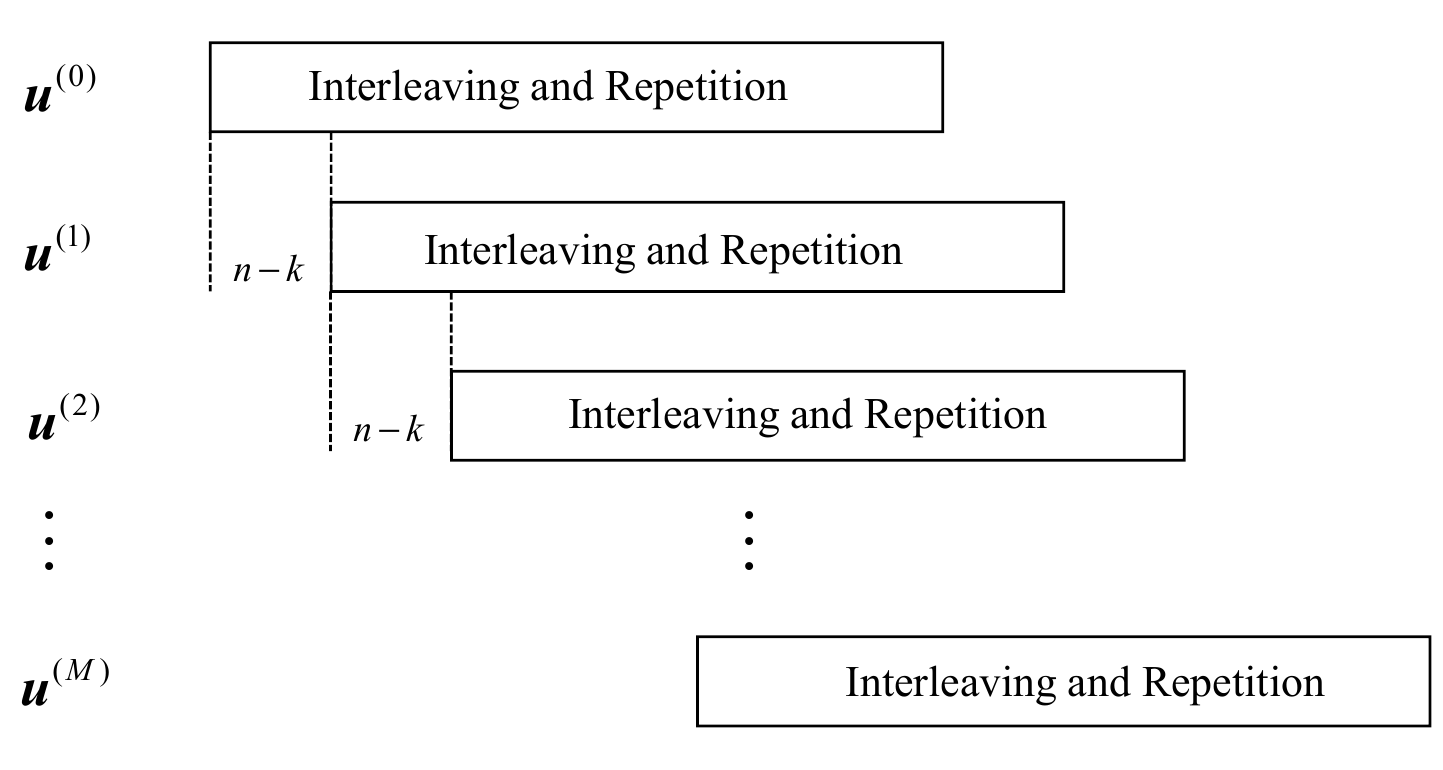}
	\caption{ A Conv-RaS code is viewed as an asynchronous multiple access system with a tunable delay, corresponding to a flexible code rate.   }
	\label{multiple_access}
\end{figure}
\subsection{Flexibility and Universality of Conv-RaS Codes}
\par To illustrate the flexibility, we treat the RaS codes as a multiple access system. 
\par \textbf{A new perspective on RaS codes:} The block RaS codes can be viewed as a multiple access system where the $m+1$ sub-blocks are treated as $m+1$ users. The ``signal'' transmitted by the $i$-th user is the interleaved replicas of the $i$-th sub-block $\bm u^{(i)}$. The receiver attempts to recover the transmitted signals from the noisy version of their binary superpositions. The Conv-RaS codes can be viewed as an asynchronous multiple access system, which has the same capacity region as the synchronous system~\cite{Cover1981asynchoronous}.  The difference between the Conv-RaS codes and the asynchronous multiple access system is that, in the Conv-RaS codes, the delay is tunable according to the code rate $k/(n-k)$, as shown in Fig.~\ref{multiple_access}. Thus, the construction of Conv-RaS codes is flexible with the code rate $k/(n-k)$ ranging from less than one ~(typically for channel codes) to greater than one~(typically for source codes), which can be implemented as a universal JSCC scheme, as confirmed by simulations. The construction of RaS codes is also universal in the sense that, without requiring complicated optimization, one pair of encoder and decoder with tunable parameters can have satisfactory performance for source coding, channel coding and JSCC. The only parameter that needs to be selected is the number of repetitions, which is directly related to the so-called genie-aided~(GA) bounds~\cite{Ma2015BMST,Cai2020SCLDGM}.
\section{Proofs of the Main Results}\label{sec4}
We need the following theorems and lemmas to proof the main results.
\subsection{Preliminary Results}
Let $P(1)=p$ and $P(0)=1-p$ be an input distribution of a BIOS memoryless channel. The mutual information between the input and the output is given by
\begin{equation}
	I(p)=(1-p)I_0(p)+pI_1(p)\text{,}
\end{equation}
where
\begin{equation}
	I_0(p)=\sum\limits_{y\in\mathcal{Y}} P(y|0)\log\frac{P(y|0)}{P(y)}\text{,}
	\label{I_{0}}
\end{equation}
\begin{equation}
	I_1(p)=\sum\limits_{y\in\mathcal{Y}} P(y|1)\log\frac{P(y|1)}{P(y)}\text{,}
\end{equation}
and $P(y)=(1-p)P(y|0)+pP(y|1)$. We define $I_0(p)$ $\big({\rm or}~I_1(p)\big)$  as \emph{partial mutual information}. For a BIOS memoryless channel, we have ${\rm max}_{0\leq p \leq 1}I(p)=I(1/2)=I_0(1/2)=I_1(1/2)$, which is the channel capacity. Notice that $I_0(p) > 0$ for $0<p<1$ as long as ${\rm Pr}\{y |P(y|0) \neq P(y|1)\} > 0$. This is a natural assumption in this paper.

\begin{lemma}
	\label{partial_mutual_information}
	The partial mutual information $I_0(p)$ is continuous, differentiable and increasing from $I_0(0)=0$ to $I_0(1)$. 
\end{lemma}
\begin{Proof}
	It can be easily seen that the partial mutual information $I_0(p)$ is continuous and differentiable for $0\leq p\leq 1$. By carrying out the differentiation, we can verify that the partial mutual information $I_0(p)$ is increasing from $I_0(0)=0$ to $I_0(1)$. 
\end{Proof}
\begin{figure}[t]
	\centering
	\includegraphics[width=0.5\textwidth]{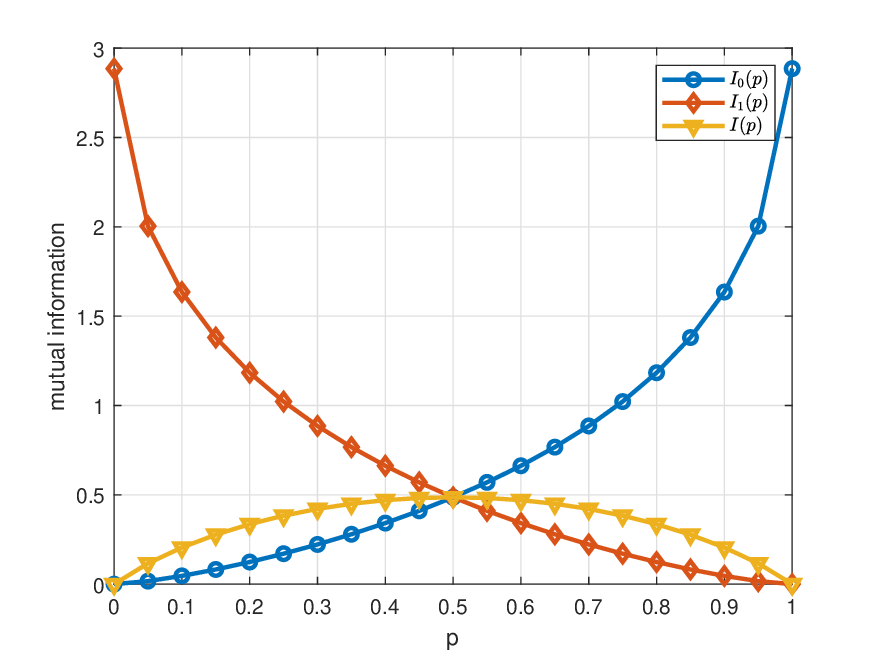}
	\caption{ The partial mutual information  $I_0(p), I_1(p)$ and the mutual information $I(p)$ with $0\leq p\leq 1$ over BI-AWGN channel with BPSK with SNR~$=0$~dB. }
	\label{MI}
\end{figure}
\begin{example}
	In Fig.~\ref{MI}, we show the partial mutual information  $I_0(p), I_1(p)$ and the mutual information $I(p)$ with~$0\leq p\leq 1$ over binary-input additive white Gaussian noise~(BI-AWGN) channel with binary phase shift keying~(BPSK) in signal-to-noise~(SNR) of~$0$~dB. The numerical results are consistent with Lemma~\ref{partial_mutual_information}.
\end{example}
\begin{lemma}\label{lemma_for_rho}
	For the considered RaS code ensemble, a message vector $\bm u^{K} =(\bm u^{(0)},\bm u^{(1)},\cdots, \\\bm u^{(m)})$ with weight profile $\bm w^{m+1}=(w_0,w_1,\cdots,w_m)$ generates a parity-check vector $\bm x \in \mathbb{F}_2^{N-K}$ as a Bernoulli sequence with success probability~\footnote{For convenience, we allow the success probability to be zero, corresponding to the degenerate Bernoulli process.}
	
	\begin{equation}
		\label{rho}
		\rho(\bm w^{m+1})=\frac{1-\prod\limits_{i=0}^{m}(1-\frac{2w_i}{k+1})}{2}\text{,}
	\end{equation}
	where $\bm w^{m+1}=(w_0,w_1,\cdots,w_m)$  is the weight profile of $\bm u^K$. Furthermore, for any non-zero message sequence $\bm u^K$, we have  $\rho(\bm w^{m+1})\geq{1}/{2}-{(k-1)}/{(k+1)}/2>0 $ and $I_0(\rho(\bm w^{m+1}))>0$.
	\par Given a positive number $T\leq m+1$, for any  given $\bm u\in \mathbb{F}_{2}^{K}$ with $t\geq T$ non-zero sub-blocks, we have
	\begin{equation}
		\begin{aligned}
			P(\bm x|\bm u)&\triangleq {\rm Pr}\{\bm X=\bm x|U=\bm u \}\leq\left[\frac{1}{2}+\frac{1}{2}\left(\frac{k-1}{k+1}\right)^{T}\right]^{N-K}
		\end{aligned}
	\end{equation}
	for all $\bm x\in \mathbb{F}_{2}^{N-K}$.
\end{lemma}
\begin{Proof}
	See Appendix.
\end{Proof}

\begin{lemma}
	\label{lemma_for_list}
	Suppose that $\bm u^K$ is transmitted over the BIOS channel and $\bm v^{K}$ is received. For any~$\delta>0$, define $A_{\delta}^{(K)}(\bm U|\bm v)$  the set of $\bm u$ sequences which are jointly $\delta$-typical  with the received $\bm v$. If $\bm v$ is typical,  then for sufficiently large $K$, the cardinality of $A_{\delta}^{(K)}(\bm U|\bm v)$ can be upper bounded by
	\begin{equation}
		\big|A_{\delta}^{(K)}(\bm U|\bm v)\big|\leq \exp\Big[K(H(U|V)+\delta)\Big].
	\end{equation}
\end{lemma}
\begin{Proof}
	See \cite[Theorem 15.2.2]{Cover2006} and it is omitted here.
\end{Proof}

\subsection{Partial Error Exponent}
In this paper, we apply the partial error
exponent to coding theorems for BIOS channels by assuming that the codeword $\bm 0$ is transmitted~\cite{Wang2022}. 
\begin{theorem}
	\label{partial_error_exponent}\rm 
	Suppose that the codeword $\bm 0\in\mathbb{F}_2^{N}$ is transmitted over a BIOS channel. Let $\mathscr{L}=\{\bm x_1,\bm x_2,\cdots,\bm x_L\}$ be a random list, where $\bm x_i\in\mathbb{F}_2^N$ is a segment of a Bernoulli process\footnote{We expect that no confusion arises from this statement that assumes only the independence among the components of $\bm x_i$. The vectors $\bm x_i$ and $\bm x_j$ can be dependent but with the identical distribution. Notice that even in the extreme case when the members in $\mathscr{L}$ are strongly correlated, say $\bm x_i = \bm x_1$ for all $i >1$, the bound for ${\rm Pr}\{ {\rm error}| \bm 0 \}$ in~\eqref{conditional_pro} still holds.} with success probability $p$. Then the probability that there exists some $i$ such that $\bm x_i$ is more likely than $\bm 0$, denoted by ${\rm Pr}\{{\rm error}|\bm 0\}$, can be upper bounded by

	\begin{equation}
		\label{conditional_pro}
		{\rm Pr}\{{\rm error}|\bm 0\}\leq \exp\Big[-NE(p,R)\Big]\text{,}
	\end{equation}
	where
	\begin{equation}
		R=\frac{1}{N}\log L\text{,}
	\end{equation}
	\begin{equation}
		E(p,R)=\max\limits_{0\leq\gamma\leq 1}(E_0(p,\gamma)-\gamma R)\text{,}
	\end{equation}
	and
	\begin{equation}
		\begin{aligned}
			E_0(p,\gamma)=-\log\Bigg\{\sum\limits_{ y\in \mathcal{Y}}P( y| 0)&^{\frac{1}{1+\gamma}}\Big[(1-p)(P( y| 0))^{\frac{1}{1+\gamma}}+p(P( y| 1))^{\frac{1}{1+\gamma}}\Big]^\gamma\Bigg\}\text{.}
		\end{aligned}	
		\label{E_independent}
	\end{equation}
	The function $E(p,R)$ is referred to as the \textit{partial error exponent}, which has the following properties:
	\begin{itemize}
		\item With a given $\gamma\in [0,1]$, $E_0(p,\gamma)$ is an increasing function of $p\in [0,1]$.
		\item With a given  $R$, $E(p,R)$ is an increasing function of  $p\in [0,1]$. 
		\item With a given $p\in [0,1]$,  $E(p,R)$ is a decreasing function of $R$ and $E(p,R)>0$ if $R<I_0(p)$.
		
	\end{itemize}
	
\end{theorem}
\begin{Proof}
Given a received sequence $\bm y$, denote by $E_i$ the event that $\bm x_i$ is more likely than $\bm 0$. For the decoding error, we have
\begin{equation}
	\begin{aligned}
		{\rm Pr}\{{\rm error}|\bm 0\}&=\sum\limits_{\bm y\in \mathcal{Y}^{N}}P(\bm y|\bm 0)\cdot {\rm {Pr}}\Big\{{\bigcup\limits_{i=1}^L E_i\Big\}}\\
		&\overset{(*)}{\leq}  L^\gamma\sum\limits_{\bm y\in \mathcal{Y}^{N}}P(\bm y|\bm 0)\bigg({\rm {Pr}}\{P(\bm y|\bm x)\geq P(\bm y|\bm 0)\}\bigg)^\gamma\text{,}
	\end{aligned}
	\label{P}	
\end{equation}
for any given $0\leq \gamma\leq 1$, where the inequality $(*)$ follows from~\cite[Lemma in Chapter 5.6]{Gallager1968}. From Markov inequality, for  $s=1/(1+\gamma)$ and a given received vector $\bm y$, the probability of a vector $\bm $ being more likely than $\bm 0$ is upper bounded by
\begin{equation}
	\begin{aligned}
		{\rm Pr}\{P(\bm y|\bm x)\geq P(\bm y|\bm 0)\}&\leq  \frac{{  \rm{\bf E}}[(P(\bm y|\bm x))^{s} ]}{(P(\bm y|\bm 0))^{s}}\\
		&=\sum\limits_{\bm  x} P(\bm x)\frac{(P(\bm y|\bm x))^{s}}{(P(\bm y|\bm 0))^{s}}\text{.}
	\end{aligned}
	\label{Markov}
\end{equation}

Substituting this bound into \eqref{P}, we have
\begin{equation}
	\begin{aligned}
		&{\rm Pr}\{{\rm error}|\bm 0\}\leq L^\gamma \sum\limits_{\bm y\in \mathcal{Y}^{N}}P(\bm y|\bm 0) \Big[\sum\limits_{\bm  x} P(\bm x)\frac{(P(\bm y|\bm x))^{s}}{(P(\bm y|\bm 0))^{s}}\Big]^\gamma\\
		&\overset{(*)}{=}L^\gamma\prod\limits_{i=0}^{N-1}\Bigg\{\sum\limits_{ y_i\in \mathcal{Y}}P(y_i|0)^{1-s\gamma}\Big[\sum\limits_{x_i\in \mathbb{F}_2} P(x_i)(P(y_i|x_i))^{s}\Big]^\gamma\Bigg\}\\
		&\overset{(**)}{\leq}\exp\Big[-NE(p,R)\Big]\text{,}
	\end{aligned}
\end{equation}
where the equality $(*)$ follows from the memoryless channel assumption and the inequality $(**)$ follows by recalling  that $s=1/(1+\gamma)$ and denoting
\begin{equation}
	E(p,R)=\max\limits_{0\leq\gamma\leq 1}(E_0(p,\gamma)-\gamma R)\text{,}
\end{equation}
and
\begin{equation}
	\begin{aligned}
		E_0(p,\gamma)=-\log\Bigg\{\sum\limits_{ y\in \mathcal{Y}}P( y| 0)^{\frac{1}{1+\gamma}}&\Big[(1-p)(P( y| 0))^{\frac{1}{1+\gamma}}+p(P( y| 1))^{\frac{1}{1+\gamma}}\Big]^\gamma\Bigg\} \text{.}
	\end{aligned}	
\end{equation}
Regarding the function $E_0(p,\gamma)$ of $p$ with a given $\gamma$, by deriving the derivation, we have 
\begin{equation}
	\begin{aligned}
		\frac{\partial E_0(p,\gamma)}{\partial p}\Bigg|_{p=0}=&-\frac{\gamma}{\ln 2}\sum\limits_{ y\in \mathcal{Y}}(P(y|0))^\frac{\gamma}{1+\gamma}\left[-(P(y|0))^{\frac{1}{1+\gamma}}+(P(y|1))^{\frac{1}{1+\gamma}}\right]\\
		&=\frac{\gamma}{\ln 2}\left[1-\sum\limits_{y \in \mathcal{Y}}P(y|0)\left(\frac{P(y|1)}{P(y|0)}\right)^{\frac{1}{1+\gamma}}\right]\text{.}
	\end{aligned}
\end{equation}
and 

\begin{equation}
	\begin{aligned}
		&	\frac{\partial^2 E_0(p,\gamma)}{\partial p^2}>0\text{.}
	\end{aligned}
\end{equation}
For a given $\gamma$ with $0\leq \gamma\leq 1$, the power function $x^{\frac{1}{1+\gamma}}$ is a concave function and we have 
\begin{equation}
	\sum\limits_{y \in \mathcal{Y}}P(y|0)\left(\frac{P(y|1)}{P(y|0)}\right)^{\frac{1}{1+\gamma}}\leq \left(\sum\limits_{y \in \mathcal{Y}}P(y|0)\frac{P(y|1)}{P(y|0)}\right)^{\frac{1}{1+\gamma}}=1
\end{equation}
since $\sum\limits_{y \in \mathcal{Y}}P(y|0)=1$ and $\sum\limits_{y \in \mathcal{Y}}P(y|1)=1$. Thus, we have 
\begin{equation}
	\begin{aligned}
		\frac{\partial E_0(p,\gamma)}{\partial p}\Bigg|_{p=0}\geq 0.
	\end{aligned}
\end{equation}
Hence, for a given $\gamma$, $E_0(p,\gamma)$ is  an increasing function of $p$.
\par Since $E_0(p, \gamma)$ is an increasing function of $p$ with a given $\gamma$, we see that $E(p,R)$ is an increasing function of $p$ with a given $R$. Obviously, for any given $0\leq p \leq 1$, $E(p,R)$ is a decreasing function of $R$.
\par Regarding the function $E_0(p,\gamma)-\gamma R$ of $\gamma$ with given $p$ and $R$, we have $E_0(p,0)-0\cdot R=0$ and
\begin{equation}
	\frac{\partial E_0(p,\gamma)}{\partial \gamma}-R\Bigg|_{\gamma=0}=I_{0}(p)-R\text{.}
\end{equation}
Hence, $E(p, R) > 0 $ if $ R < I_0(p)$. 
\end{Proof}

\par \textbf{Remarks:} For the partial error exponent, we have the following remarks:
\begin{itemize}
	\item From the above proof, we see that the members in the list need to have the same
	distribution or  have the same upper bounds of probability but the condition of pair-wise independence is not necessary, which is distinguished from the proof in~\cite[Chapter 5]{Gallager1968}.
	\item   Partial error exponent derived in this paper applies only to BIOS channels while the error exponent derived in~\cite{Gallager1968} can apply to any stationary and memoryless channel.
\end{itemize}

\subsection{Proof of Coding Theorem for RaS Code Ensemble}
\label{proof_block}
\textit{Proof of Theorem~\ref{theorem_for_block_code}:}	Suppose that $(\bm u^{K},\bm u^{K}\mathbf{G})$ is transmitted. Let $\epsilon>0$ and $\delta>0$ be two arbitrarily small numbers. Upon receiving $(\bm v^{K},\bm y^{N-K})$, we use the following two-step decoding. First, list all sequences $\bm {\tilde u}$ such that $(\bm {\tilde{u}},\bm v^{K})$  are jointly $\delta$-typical. Second, find from the list a sequence $\hat{\bm u}$ such that $P(\bm y|\hat{\bm u}\mathbf{G})$ is maximized.
	\par There are two types of errors. One is the case when $(\bm {u}^K,\bm v^{K})$ are not jointly $\delta$-typical and hence $\bm u^{K}$ is not in the list. This type of error, from \cite[Chapter 7]{Cover2006}, can have an arbitrarily small probability as long as $m$~(hence $K$) is sufficiently large.
	\par The other case is that $\bm u^{K}$ is in the list but is not the most likely one.  In this case, denote the list as ${\mathscr{\tilde{L}}}=\{{\bm \tilde{\bm u}}_0=\bm u^{K},{\bm \tilde{\bm u}}_1,\cdots,{ \tilde{\bm u}}_L\}$. From Lemma~\ref{lemma_for_list}, we have~$L\leq \exp\left[K(H(U|V)+\delta)\right]$ for sufficiently large $K$. Given the received sequence $\bm y$ and the list  ${\mathscr{L}}$, the decoding output~$\hat{\bm U}$ is a random sequence over the code ensemble due to the randomness of $\mathbf{G}$.  Given a received sequence $\bm y$, denote by~$E_{\bm u^K\rightarrow \tilde{\bm u}_i}$ the event that $\tilde{\bm u}_i\mathbf{G}$ is more likely than $\bm u^K\mathbf{G}$.  We have
	\begin{equation}
		\label{error_uk}
		\begin{aligned}
			{\rm{Pr}}\{{\rm error}|\bm u^{K}\}
			&\leq \frac{\epsilon}{3}+\sum\limits_{ \bm x\in \mathbb{F}_2^{N-K}}\sum\limits_{\bm y \in \mathcal{Y}^{N-K}} P(\bm u^K\mathbf{G}=\bm x) P(\bm y|\bm x)\cdot {\rm {Pr}}\Bigg\{{\bigcup\limits_{i=1}^LE_{\bm u^K\rightarrow \tilde{\bm u}_i}\Bigg\}}.
		\end{aligned}	
	\end{equation}
Given $\bm y\in \mathcal{Y}^{N-K}$, $\bm x\in\mathbb{F}_2^{N-K}$, we define
\begin{equation}
	\label{transform_for_0}
	\pi^{x_i}(y_i)=
	\begin{cases}
		y_{i}& {\rm if~} x_i=0\\
		\pi(y_i)& {\rm if~}x_i=1
	\end{cases}.
\end{equation}
Then we have $P(\bm y|\bm x)=P(\pi^{\bm x}(\bm y)|\bm 0)$ and $P(\bm y|\tilde{\bm u}_i\mathbf{G})\geq P(\bm y|{\bm u^K}\mathbf{G} )$ is equivalent to $P(\pi^{\bm x}(\bm y)|(\tilde{\bm u}_i+\bm u^K)\mathbf{G})\geq P(\pi^{\bm x}(\bm y)|\bm 0)$. Therefore, the performance can be analyzed by assuming that $\bm 0$ is transmitted and competed with a list of codewords $\bm u_i\mathbf{G}$ at the decoder, where $\bm u_i=\tilde{\bm u}_i+\bm u^K$ for $1\leq i\leq L$. With this equivalence, we have
\begin{equation}
	\label{error0}
	\begin{aligned}
			{\rm{Pr}}\{{\rm error}|\bm u^{K}\}
			&\leq \frac{\epsilon}{3}+\sum\limits_{ \bm x\in \mathbb{F}_2^{N-K}}\sum\limits_{\bm y \in \mathcal{Y}^{N-K}} P(\bm u^K\mathbf{G}=\bm x) P(\bm y|\bm x)\cdot {\rm {Pr}}\Bigg\{{\bigcup\limits_{i=1}^LE_{\bm u^K\rightarrow \tilde{\bm u}_i}\Bigg\}}\\
		&=\frac{\epsilon}{3}+\sum\limits_{\bm y\in \mathcal{Y}^{N-K}}P(\bm y|\bm 0)\cdot {\rm {Pr}}\Big\{{\bigcup\limits_{i=1}^LE_i\Big\}}\text{,}
	\end{aligned}	
\end{equation}
where $E_i$ is the event that, given a received sequence $\bm y$, $\bm u_i\mathbf{G}$ is more likely than $\bm 0$.
\par    We partion the list $\mathscr{L}=\{\bm u_0=\bm 0,\bm u_1,\cdots,\bm u_L\}$ according to the weight profile $\bm w$. For a weight profile $\bm w\in [0:k]^{m+1}$, denote by~$\mathscr{L}_{\bm w}$ the set of all sequences $\bm u=(\bm u^{(0)},\bm u^{(1)},\cdots,\bm u^{(m)})\in \mathscr{L}$ having the weight profile $\bm w$.  We have
\begin{equation}
	\mathscr{L}=\bigcup\limits_{\bm w\in [0:k]^{m+1}}\mathscr{L}_{\bm w}\text{.}
\end{equation}
\par Let $T=\lfloor \sqrt{m+1}\rfloor$, the maximum integer such that $T^2\leq m+1$. The error event given $\bm y$ can be split into two sub-events depending on whether the Hamming weight $W_H(\bm w)$~(the number of non-zero components of $\bm w$)  is greater than $T$ or not. We have
\begin{equation}
	\begin{aligned}
		\sum\limits_{\bm y\in \mathcal{Y}^{N-K}}P(\bm y|\bm 0){\rm {Pr}}\Bigg\{{\bigcup\limits_{i=1}^LE_i\Bigg\}}
		&\leq\sum\limits_{\bm y\in \mathcal{Y}^{N-K}}P(\bm y|\bm 0)\sum\limits_{\substack{\bm w\in [0:k]^{m+1}\\1\leq W_H(\bm w)\leq T}}{\rm Pr}\left\{\bigcup\limits_{ \bm u\in\mathscr{L}_{\bm w}} P(\bm y|\bm u\mathbf{G})\geq P(\bm y|\bm 0)\right\}\\
		&+\sum\limits_{\bm y\in \mathcal{Y}^{N-K}}P(\bm y|\bm 0)~{\rm Pr}\left\{\bigcup\limits_{\substack{\bm w\in [0:k]^{m+1}\\ W_H(\bm w)> T}}\bigcup\limits_{ \bm u\in\mathscr{L}_{\bm w}} P(\bm y|\bm u\mathbf{G})\geq P(\bm y|\bm 0)\right\}\text{.}
	\end{aligned}
	\label{bound1}
\end{equation}
\par   For a non-zero weight profile~$\bm w=({w}_0,{w}_1,\cdots,{w}_{m})\in [0:k]^{m+1}$, we have
\begin{equation}
	\begin{aligned}
	&\sum\limits_{\bm y\in \mathcal{Y}^{N-K}}P(\bm y|\bm 0){\rm Pr}\left\{\bigcup\limits_{ \bm u\in\mathscr{L}_{\bm w}} P(\bm y|\bm u\mathbf{G})\geq P(\bm y|\bm 0)\right\}\\
&\overset{(*)}{\leq} \exp\left[{-(N-K) E\left(\rho(\bm w),{{R}}(\bm w)\right)}\right]\\
&\overset{(**)}{\leq }	\exp\left[-(N-K)E\left(\rho_1,R_1\right)\right]\text{,}
\end{aligned}
\end{equation}
where the inequality $(*)$ follows from the proof of Theorem~\ref{partial_error_exponent} by denoting
\begin{equation}
	\label{lower}
\rho(\bm w)=\frac{1-\prod\limits_{i=0}^{m}(1-\frac{2w_i}{k+1})}{2}\text{,}
\end{equation}   
\begin{equation}
	\label{upper}
	{R}(\bm w)=\frac{\log\left[\prod_{i=0}^{m}\binom{k}{{w}_i}\right]}{N-K}
\end{equation}
and the inequality~$(**)$ follows from Theorem~\ref{partial_error_exponent} by taking into account the lower bound $\rho(\bm w)\geq\rho_1\triangleq{1}/{2}-({k-1})/(k+1)/2$ from Lemma~\ref{lemma_for_rho} and the upper bound $R(\bm w)\leq R_1\triangleq kT/(N-K)$.
\par Since the number of non-zero $\bm w$ with $W_H(\bm w)\leq T$ is upper bounded by $(m+1)\binom{m+1}{T}(k+1)^{T}$, we have  
\begin{equation}
	\label{error_leqT}
	\begin{aligned}
			&\sum\limits_{\bm y\in \mathcal{Y}^{N-K}}P(\bm y|\bm 0)\sum\limits_{\substack{\bm w\in [0:k]^{m+1}\\1\leq W_H(\bm w)\leq T}}{\rm Pr}\left\{\bigcup\limits_{ \bm u\in\mathscr{L}_{\bm w}} P(\bm y|\bm u\mathbf{G})\geq P(\bm y|\bm 0)\right\}\\
			&{\leq} (m+1)\binom{m+1}{T}(k+1)^{T}\exp\left[-(N-K)E\left(\rho_1,R_1\right)\right]\\
			&\overset{(*)}{\leq } \exp\left\{-(m+1)\left[(n-k)E\left(\rho_1,R_1\right)-\frac{\log (m+1) }{m+1}-H\left(\frac{T}{m+1}\right)-\frac{T\log(k+1)}{m+1}\right]\right\}\text{,}
	\end{aligned}
\end{equation}
where the inequality $(*)$ follows from the fact that $\binom{m+1}{T}\leq \exp\left[(m+1)H(T/(m+1))\right]$ with~$H(\cdot)$ being the entropy function~\cite[Example 11.1.3]{Cover2006}.
\par For $W_H(\bm w)> T$, we have, for any $0\leq \gamma \leq 1$,  
\begin{equation}
	\begin{aligned}	
		&\sum\limits_{\bm y\in \mathcal{Y}^{N-K}}P(\bm y|\bm 0){\rm Pr}\left\{\bigcup\limits_{\substack{\bm w\in [0:k]^{m+1}\\ W_H(\bm w)> T}}\bigcup\limits_{ \bm u\in\mathscr{L}_{\bm w}} P(\bm y|\bm u\mathbf{G})\geq P(\bm y|\bm 0)\right\}\\
		&\leq \sum\limits_{\bm y\in \mathcal{Y}^{(N-K)}}P(\bm y|\bm 0)\left(\sum\limits_{\substack{\bm w \in[0:k]^{m+1}\\W_H(\bm w)>T\\\bm u \in \mathscr{L}_{\bm w}}}{\rm Pr}\{P(\bm y|\bm u\mathbf{G})\geq P(\bm y|\bm 0)\}\right)^{\gamma}\\	
		&\leq \sum\limits_{\bm y\in \mathcal{Y}^{N-K}}P(\bm y|\bm 0)\left\{|\mathscr{L}|\left[\frac{1}{2}+\frac{1}{2}\left(\frac{k-1}{k+1}\right)^{T}\right]^{N-K}\sum\limits_{\bm x \in \mathbb{F}_2^{N-K}} \frac{(P(\bm y|\bm x))^{s}}{(P(\bm y|\bm 0))^{s}}\right\}^{\gamma}\\
		&\overset{(*)}{\leq}\exp\left[{K\gamma (H(U|V)+\delta)}\right]\left[1+\left(\frac{k-1}{k+1}\right)^{T}\right]^{(N-K)\gamma}\\
		&\qquad\qquad\cdot \left[\sum\limits_{ y_i\in \mathcal{Y}}(P(y_i|0))^{1-s\gamma}\left(\sum\limits_{ x_i\in \mathbb{F}_2}\frac{1}{2}\cdot{(P( y_{i}|x_{i}))^{s}}\right)^{\gamma}\right]^{N-K}\\
		&\overset{(**)}{\leq} \exp\left[{-(N-K) E\left(\frac{1}{2},{R}_T\right)}\right]\text{,}
	\end{aligned}
	\label{latterbound}
\end{equation}
where the inequality $(*)$ follows from  the  BIOS memoryless channel assumption and the upper bound for $|\mathscr{L}|$ with sufficiently large $K$, and the inequality $(**)$ follows from Theorem~\ref{partial_error_exponent} by denoting

\begin{equation}
	{R}_T= \log\left[1+\left(\frac{k-1}{k+1}\right)^T\right]+ \frac{k}{n-k} (H(U|V)+\delta)\text{.}
\end{equation}

\par Combining~\eqref{error0},~\eqref{bound1},~\eqref{error_leqT} and~\eqref{latterbound}, we have
\begin{equation}
	\label{result}
	\begin{aligned}
		&{\rm{Pr}}\{{\rm error}|\bm u^{K}\}\leq \frac{\epsilon}{3}+\exp\left[{-(N-K)E\left(\frac{1}{2},{R}_T\right)}\right]\\
		&+\exp\left\{-(m+1)\left[(n-k)E\left(\rho_1,R_1\right)-\frac{\log (m+1) }{m+1}-H\left(\frac{T}{m+1}\right)-\frac{T\log(k+1)}{m+1}\right]\right\}\text{.}		
	\end{aligned}
\end{equation}
\par For  $k/(n-k)<I(X;Y)/H(U|V)$, letting $m \rightarrow \infty$ (and hence $T \rightarrow \infty$),  we have, for sufficiently small $\delta$, ${R}_T \rightarrow(H(U|V)+\delta)k/(n-k)<I(X;Y)\leq C$, where $C$ is the capacity of BIOS~2. Thus, we have from Theorem~\ref{partial_error_exponent} that  $E(1/2, {R}_T) > 0 $ for $R_T<I_0(1/2)$. The second term in the right hand side~(RHS) of the inequality~\eqref{result} can be made not greater than $\epsilon/3$. For $m \rightarrow \infty$, we have $R_1\rightarrow0$, $\log (m+1)/(m+1)\rightarrow0$, $H(T/(m+1))\rightarrow0$ and $T\log(k+1)/(m+1)\rightarrow0$. From Lemma~\ref{lemma_for_rho}, we have $I_0(\rho_1)>I_0(0)=0$.  Hence, we have $E(\rho_1,R_1)>0$ from Theorem~\ref{partial_error_exponent}. The third term in the RHS of the inequality~\eqref{result} can be made not greater than $\epsilon/3$.

Now we have
\begin{equation}
	{\rm Pr}\{{\rm error}|\bm u^K\} \leq \epsilon.
\end{equation}
Therefore, ${\rm Pr}\{{\rm error} \} = \sum_{\bm u^{K} \in \mathbb{F}_2^{K} } 2^{-K} {\rm Pr}\{{\rm error}|\bm u^K\} \leq  \epsilon$.
\par This completes the proof of Theorem~\ref{theorem_for_block_code}.
\par {\textbf{Remarks:}} From the proof, we see that the systematic bits and the parity-check bits play different roles. Receiving noisy systematic bits provides us with a list of the source output while receiving noisy parity-check bits helps us to select the correct one from the list. We also see that, for the RaS code ensemble to have an arbitrarily small error probability, the number of sub-blocks~$(m+1)$ should be sufficiently large while the length of sub-blocks~$k$~(hence $n$) can be any fixed positive integers as long as $k/(n-k)<I(X;Y)/H(U|V)$. This indicates that $k$ and $n$ can be chosen properly such that the gap between the ratio $k/(n-k)$ and the corresponding limit~$I(X; Y)/H(U|V)$ is as small as desired. However, for large $m$, the MLD is usually impractical. For this reason, we turn to the iterative belief propagation~(BP) decoding, which is effective for sparse codes, in which case relatively large $k$~(hence $n$) is preferable.

\subsection{Proof of Coding Theorem for Conv-RaS Code Ensemble}
\label{proof_SC}
 \par\textit{Proof of Theorem~\ref{theorem_for_conv}:} At the receiver, we consider a suboptimal decoding~\cite[Page 417]{jacobs1965principles} algorithm which takes as input the truncated vector $\tilde{\bm y}=(\bm y^{(0)},\bm y^{(1)},\cdots,\bm y^{(m)})$ and delivers as output an estimate of $\bm u^{(0)}$, denoted by $\hat{\bm u}^{(0)}$.  In this case, the truncated generator matrix is 
 \begin{equation}
 	\label{trancated}
 	\mathbf{G} = \left[
 	\begin{array}{cccccccc}
 		\mathbf{G}_{0,0} & \mathbf{G}_{0,1} &  \cdots    & \mathbf{G}_{0,m}\\
 		& \mathbf{G}_{1,1} & \cdots  & \mathbf{G}_{1,m}\\
 		&         & \ddots  & \vdots   \\
 		&         &  & \mathbf{G}_{m,m}     \\
 	\end{array}
 	\right]\text{.}
 \end{equation} 
 \par Consider the message vector $\bm u^{m+1}=(\bm u^{(0)},\bm u^{(1)},\cdots,\bm u^{(m)})\in \mathbb{F}_2^{K}$ of weight profile~$\bm w^{m+1}=(w_0,w_1,\cdots
	,w_m)$ with $W_H(\bm u^{(i)})=w_i$ for $0\leq i\leq m$. Suppose that $\bm u^{K}\mathbf{G}$ is transmitted. Let~$\epsilon>0$ and $\delta>0$ be two arbitrarily small numbers. Upon receiving $\bm y^{N-K}$, we use the following two-step decoding. First, list all sequences $\tilde{\bm u}$  with
	\begin{equation}
		\mathscr{L}=\{\tilde{\bm u}\in \mathbb{F}_2^K: H(P_{\tilde{\bm u}})\leq H(U)+\delta\},
	\end{equation}
	where $P_{\tilde{\bm u}}$ is the type of $\tilde{\bm u}$~\cite[Chapter 11]{Cover2006}. Second, find from $\mathscr{L}$ a sequence $\hat{\bm u}$ such that~$P(\bm y|\hat{\bm u}\mathbf{G})$ is maximized.
	\par There are two types of errors. One is the case when $\bm u^{K}$ is not in the list. This type of errors, from \cite[Chapter 11]{Cover2006}, have a probability going to $0$ exponentially with  sufficiently large~$m $~(hence $K$). For sufficiently large $K$, the list size can be upper bounded by $L\leq \exp[K(H(U)+\delta)]$. Actually, for sufficiently large $K^\prime<K$, we have $|\mathscr{L}^\prime|\leq \exp[K^\prime(H(U)+\delta)]$ where
	\begin{equation}
		\mathscr{L}^\prime=\{\tilde{\bm u}^\prime\in \mathbb{F}_2^{K^\prime}:~\tilde{\bm u}^\prime~\text{is obtained by discarding the first}~K-K^\prime~\text{components of some}~\tilde{\bm u} \in\mathscr{L}~\}.
	\end{equation}
	
	We can use the transform in~\eqref{transform_for_0} and consider the case that the all zero message vector is transmitted. Thus, we have
	\begin{equation}
		\label{SC_error}
		\begin{aligned}
			{\rm{Pr}}\{{\rm error}|\bm u^{K}\}
			&\leq \frac{\epsilon}{3}+\sum\limits_{ \bm x\in \mathbb{F}_2^{N-K}}\sum\limits_{\bm y \in \mathcal{Y}^{N-K}} P(\bm u^K\mathbf{G}=\bm x) P(\bm y|\bm x)\cdot {\rm {Pr}}\Bigg\{{\bigcup\limits_{i=1}^LE_{\bm u^K\rightarrow \tilde{\bm u}_i}\Bigg\}}\\
			&=\frac{\epsilon}{3}+\sum\limits_{\bm y\in \mathcal{Y}^{N-K}}P(\bm y|\bm 0) {\rm {Pr}}\Big\{{\bigcup\limits_{i=1}^LE_i\Big\}}\text{,}
		\end{aligned}	
	\end{equation}
	where $E_i$ is the event that, given a received sequence $\bm y$ associated with the transmitted sequence~$\bm 0$, $\bm u_i\mathbf{G}$ is more likely than $\bm 0$.
	%
	\par We partion the list $\mathscr{L}=\{\bm u_0=\bm 0,\bm u_1,\cdots,\bm u_L\}$ according to the weight profile $\bm w$. For a weight profile $\bm w\in [0:k]^{m+1}$, denote by~$\mathscr{L}_{\bm w}$ the set of all sequences $\bm u=(\bm u^{(0)},\bm u^{(1)},\cdots,\bm u^{(m)})\in \mathscr{L}$ having the weight profile $\bm w$.  We have
	\begin{equation}
		\mathscr{L}=\bigcup\limits_{\bm w\in [0:k]^{m+1}}\mathscr{L}_{\bm w}\text{.}
	\end{equation}
	
	\par Let $T=\lfloor \sqrt{m+1}\rfloor$. The error event given $\bm y$ can be split into two sub-events depending on whether the Hamming weight $W_H(\bm w)$~(the number of non-zero components of $\bm w$)  is greater than $2T$ or not. We have
	
	\begin{equation}
		\begin{aligned}
			\sum\limits_{\bm y\in \mathcal{Y}^{N-K}}P(\bm y|\bm 0){\rm {Pr}}\Bigg\{{\bigcup\limits_{i=1}^LE_i\Bigg\}}
			&\leq\sum\limits_{\bm y\in \mathcal{Y}^{N-K}}P(\bm y|\bm 0)\sum\limits_{\substack{\bm w\in [0:k]^{m+1}\\1\leq W_H(\bm w)\leq 2T}}{\rm Pr}\left\{\bigcup\limits_{ \bm u\in\mathscr{L}_{\bm w}} P(\bm y|\bm u\mathbf{G})\geq P(\bm y|\bm 0)\right\}\\
			&+\sum\limits_{\bm y\in \mathcal{Y}^{N-K}}P(\bm y|\bm 0)~{\rm Pr}\left\{\bigcup\limits_{\substack{\bm w\in [0:k]^{m+1}\\ W_H(\bm w)> 2T}}\bigcup\limits_{ \bm u\in\mathscr{L}_{\bm w}} P(\bm y|\bm u\mathbf{G})\geq P(\bm y|\bm 0)\right\}\text{.}
		\end{aligned}
		\label{SC_partion}
	\end{equation}
	
	\par For $W_H(\bm w)\leq 2T$, similar to the proof in Theorem~\ref{theorem_for_block_code}, we have, for any $0\leq \gamma\leq 1$
	
	\begin{equation}
		\label{sc_partion_1}
		\begin{aligned}
			&\sum\limits_{\bm y\in \mathcal{Y}^{N-K}}P(\bm y|\bm 0))\sum\limits_{\substack{\bm w\in [0:k]^{m+1}\\1\leq W_H(\bm w)\leq 2T}}{\rm Pr}\left\{\bigcup\limits_{\bm u\in \mathscr{L}_{\bm w}}P(\bm y|\bm u\mathbf{G})\geq P(\bm y|\bm 0)\right\}\\
			&\overset{(*)}{\leq} \sum\limits_{\bm y\in \mathcal{Y}^{N-K}}P(\bm y|\bm 0)\sum\limits_{\substack{\bm w\in [0:k]^{m+1}\\1\leq W_H(\bm w)\leq 2T}}\left(\exp(2Tk)\sum\limits_{ \bm x\in \mathbb{F}_2^{N-K}}P(\bm x)\left(\frac{P(\bm y|\bm x)}{P(\bm y|\bm 0)}\right)^s\right)^\gamma\\
			&\overset{(**)}{=}\sum\limits_{\substack{\bm w\in [0:k]^{m+1}\\1\leq W_H(\bm w)\leq 2T}}\exp\left\{-(N-K)\left[\frac{1}{m+1}E_0(\rho(\bm w^1),\gamma)+\cdots+\frac{1}{m+1}E_0(\rho(\bm w^{m+1}),\gamma)-\frac{\gamma 2Tk}{N-K}\right]\right\}\\
			&\overset{(***)}{\leq}(m+1)\binom{m+1}{2T}(k+1)^{2T}\exp\left[-(N-K)E\left(\rho_1,R_2\right)\right]\\
			&\overset{(****)}{\leq} \exp\left\{-(m+1)\left[(n-k)E(\rho_1,R_2)-\frac{\log(m+1)}{m+1}-H\left(\frac{2T}{m+1}\right)-\frac{2T\log(k+1)}{m+1}\right]\right\}\text{,}
		\end{aligned}
	\end{equation}
	where the inequality $(*)$ follows from the fact that the number of sequences $\bm u$ with a given $\bm w~(W_H(\bm w)\leq 2T)$ is upper bounded by $\exp(2Tk)$, the equality $(**)$ follows from the proof of  Lemma~\ref{lemma_for_rho} and Theorem~\ref{partial_error_exponent} by dealing  separately with each sub-block $\bm u^{(i)}$, the inequality~$(***)$ follows  from the number of non-zero $\bm w$ with $W_H(\bm w)\leq2 T$ is upper bounded by $(m+1)\binom{m+1}{2T}(k+1)^{2T}$ and Theorem~\ref{partial_error_exponent} by denoting
	\begin{equation}
		\rho_1=\frac{1}{2}-\frac{k-1}{2(k+1)}
	\end{equation}
	and
	\begin{equation}
		R_2=\frac{2Tk}{N-K}\text{,}
	\end{equation}
	and the inequality~$(****)$ follows from  the fact that $\binom{m+1}{2T}\leq \exp[(m+1)H\left(\frac{2T}{m+1}\right)]$ with $H(\cdot)$ being the entropy function~\cite[Example 11.1.3]{Cover2006}.
	
	For a weight profile $\bm w\in [0:k]^{m+1}$ with $W_H(\bm w)>2T$, we define
	\begin{equation}
		\tau(\bm w)=\min\{t\geq0|~(w_0,w_1,\cdots,w_t)~\text{has}~\text{exactly}~T~\text{non-zero elements}\}\text{,}
	\end{equation}
	where the sequence $(w_0,w_1,\cdots,w_t)$ is the first $t+1$ components of $\bm w^{m+1}$. For any given $\bm h\in \mathbb{F}_2^{k(\tau(\bm w)+1)}$, we define 
	\begin{equation}
		\mathscr{L}_{\tau(\bm w), \bm h}=\{\bm u\in \mathscr{L} : \bm u^{k(\tau(\bm w)+1)}=\bm h^{k(\tau(\bm w)+1)}\}.
	\end{equation}
	Thus, for $W_H(\bm u)>2T$ and $T-1\leq \tau(\bm w)\leq m-T$, we have 
	\begin{equation}
		\bigcup\limits_{\substack{\bm w\in [0:k]^{m+1}\\ W_H(\bm w)>2T}}\mathscr{L}_{\bm w}=\bigcup\limits_{ T-1\leq \tau(\bm w)\leq m-T}\bigcup\limits_{\bm h\in \mathbb{F}_2^{k(\tau(\bm w)+1)}}\mathscr{L}_{\tau(\bm w),\bm h}.
	\end{equation}
	For a given $\tau(\bm w)$ such that $T-1\leq \tau(\bm w)\leq m-T$ and a given $\bm h$, we have 
	\begin{equation}
		\begin{aligned}	
			& \sum\limits_{\bm y\in \mathcal{Y}^{N-K}}P(\bm y|\bm 0) {\rm Pr}\left\{\bigcup\limits_{\bm u\in \mathscr{L}_{\tau(\bm w),\bm h}}P(\bm y|\bm u\mathbf{G})\geq P(\bm y|\bm 0)\right\}\\	
			&\overset{(*)}{\leq} \sum\limits_{\bm y\in \mathcal{Y}^{N-K}}P(\bm y|\bm 0)\bigg\{ \sum\limits_{ \bm x\in \mathbb{F}_2^{(\tau(\bm w)+1)(n-k)}}P(\bm x)\left(\frac{P(\bm y|\bm x)}{P(\bm y|\bm 0)}\right)^s\left[\frac{1}{2}+\frac{1}{2}\left(\frac{k-1}{k+1}\right)^{T}\right]^{(m-\tau(\bm w))(n-k)}\\
			&\qquad\qquad\exp\left[k(m-\tau(\bm w))(H(U)+\delta)\right]\sum\limits_{ \bm x\in \mathbb{F}_2^{(m-\tau(\bm w))(n-k)}}\left(\frac{P(\bm y|\bm x)}{P(\bm y|\bm 0)}\right)^s\bigg\}^\gamma\\
			&\overset{(**)}{\leq} \exp\bigg\{-(N-K) \bigg[\frac{\tau(
				\bm w)+1}{m+1}E_0\left(\frac{1}{2}-\frac{k-1}{2(k+1)},\gamma\right)+\frac{m-\tau(\bm w)}{m+1}E_0\left(\frac{1}{2},\gamma\right)\\
			&-\frac{\gamma k(m-\tau(\bm w))(H(U)+\delta)}{(m+1)(n-k)}-\frac{\gamma (m-\tau(\bm w))\log\left[1+\left(\frac{k-1}{k+1}\right)^T\right]}{m+1}\bigg]\bigg\},\\
		\end{aligned}
	\end{equation}
	where the inequality $(*)$ follows from that the number of sequences is upper bounded by $\exp[k(m-\tau(\bm w))(H(U)+\delta)]$, by dealing with each sub-block in $\bm u$ and from Lemma~\ref{lemma_for_rho} and the inequality~$(**)$ follows from Lemma~\ref{lemma_for_rho} and Theorem~\ref{partial_error_exponent}.
	\par Since the number of the sequences $\bm h$ for a given $\tau(\bm w)$ is upper bounded by $ \binom{m+1}{T} \exp(kT)$, we have
	
	\begin{equation*}
		\begin{aligned}
			&\sum\limits_{\bm y\in \mathcal{Y}^{N-K}}P(\bm y|\bm 0){\rm Pr}\left\{\bigcup\limits_{\substack{\bm w\in [0:k]^{m+1}\\ W_H(\bm w)> 2T}}\bigcup\limits_{\bm u\in \mathscr{L}_{\bm w}}P(\bm y|\bm u\mathbf{G})\geq P(\bm y|\bm 0)\right\}\\
			&= \sum\limits_{\bm y\in \mathcal{Y}^{N-K}}P(\bm y|\bm 0){\rm Pr}\left\{\bigcup\limits_{ T-1\leq \tau(\bm w)\leq m-T}\bigcup\limits_{\bm h\in \mathbb{F}_2^{k(\tau(\bm w)+1)}}\bigcup\limits_{\bm u\in \mathscr{L}_{\tau(\bm w),\bm h}}P(\bm y|\bm u\mathbf{G})\geq P(\bm y|\bm 0)\right\}\\
			&\leq \sum\limits_{\tau(\bm w)=T-1}^{m-T}\sum\limits_{\bm h\in \mathbb{F}_2^{k(\tau(\bm w)+1)}}\sum\limits_{\bm y\in \mathcal{Y}^{N-K}}P(\bm y|\bm 0){\rm Pr}\left\{\bigcup\limits_{\bm u\in \mathscr{L}_{\tau(\bm w),\bm h}}P(\bm y|\bm u\mathbf{G})\geq P(\bm y|\bm 0)\right\}\\
				&\leq\sum\limits_{\tau(\bm w)=T-1}^{m-T} \binom{m+1}{T} \exp(kT)\exp\bigg\{-(N-K) \bigg[\frac{\tau(\bm w)+1}{m+1}E_0\left(\frac{1}{2}-\frac{k-1}{2(k+1)},\gamma\right)\qquad\qquad\qquad
		\end{aligned}
	\end{equation*}
\begin{equation}
	\label{sc_2}
	\begin{aligned}
			&+\frac{m-\tau(\bm w)}{m+1}E_0\left(\frac{1}{2},\gamma\right)-\frac{\gamma k(m-\tau(\bm w))(H(U)+\delta)}{(m+1)(n-k)}-\frac{\gamma (m-\tau(\bm w))\log\left[1+\left(\frac{k-1}{k+1}\right)^T\right]}{m+1}\bigg]\bigg\}	\\
		&\overset{(*)}{\leq}\sum\limits_{\tau(\bm w)=T-1}^{m-T}\exp\left\{-(m+1)\left[(n-k)E(\tau(\bm w),R_3)-H\left(\frac{T}{m+1}\right)-\frac{kT}{m+1}\right]\right\}\text{,}\\
	\end{aligned}
\end{equation}
	where  the inequality $(*)$ follows from~\cite[Example 11.1.3]{Cover2006} and from the proof in Theorem~\ref{partial_error_exponent} by denoting 
	\begin{equation}
		E(\tau(\bm w),R_3)=\max\limits_{0\leq \gamma \leq 1}\left[\frac{\tau(\bm w)+1}{m+1}E_0\left(\rho_1,\gamma\right)+\frac{m-\tau(\bm w)}{m+1}\left(E_0\left(\frac{1}{2},\gamma\right)-\gamma R_3\right)\right]
		\text{,}
	\end{equation}
	and 
	\begin{equation}
		R_3=\frac{ k(H(U)+\delta)}{n-k}+ \log\left[1+\left(\frac{k-1}{k+1}\right)^T\right].
	\end{equation}
	\par Combining~\eqref{SC_error},~\eqref{sc_partion_1} and \eqref{sc_2}, we have
	\begin{equation}
		\begin{aligned}
			&{\rm{Pr}}\{{\rm error}|\bm u^{K}\}\leq \frac{\epsilon}{3}+\sum\limits_{\tau(\bm w)=T-1}^{m-T}\exp\left\{-(m+1)\left[(n-k)E(\tau(\bm w),R_3)-H\left(\frac{T}{m+1}\right)-\frac{kT}{m+1}\right]\right\}\\
			&+\exp\left\{-(m+1)\left[(n-k)E(\rho_1,R_2)-\frac{\log(m+1)}{m+1}-H\left(\frac{2T}{m+1}\right)-\frac{2T\log(k+1)}{m+1}\right]\right\}
			\text{.}
			\label{sc_result}
		\end{aligned}
	\end{equation}
	\par    Letting $m\rightarrow\infty$~(hence $T\rightarrow\infty$), since $1+\left(\frac{k-1}{k+1}\right)^T \rightarrow1$ and  $k/(n-k)<I(X;Y)/H(U)$, we have, for sufficiently small $\delta$, ${R}_3\rightarrow k(H(U)+\delta)/(n-k) < I(X; Y) \leq C$, where $C$ is the capacity of BIOS~2. Since $I_0(\rho_1)>I_0(0)=0$, we have from Theorem~\ref{partial_error_exponent} that  $E(\tau(\bm w), {R}_3) > 0 $ for~$R_3<I_0(1/2)$.  We also have  $H(T/(m+1))\rightarrow0$ and $kT/(m+1)\rightarrow 0$ for $m\rightarrow\infty$. Thus, for a given $\tau(\bm w)$, one term of the second summation in the RHS of the inequality~\eqref{sc_result} goes to $0$ exponentially with $m$. Since the number of terms is less than $m+1$, the second term of the RHS of the inequality of~\eqref{sc_result} can be made not greater than $\epsilon/3$.  For $m \rightarrow\infty$, we have, $H(2T/(m+1))\rightarrow0$, ${\log(m+1)}/{(m+1)}\rightarrow 0$, $2T\log(k+1)/(m+1)\rightarrow 0$ and $R_2\rightarrow 0$. Since $I_0(\rho_1)>0$ from Lemma~\ref{lemma_for_rho},  we have  $E(\rho_1,R_2)>0$ from Theorem~\ref{partial_error_exponent}. The third term of the RHS of the inequality~\eqref{sc_result} can be not greater than $\epsilon/3$.
	
	Now we have
	\begin{equation}
		{\rm Pr}\{{\rm error}|\bm u^K\} \leq \epsilon.
	\end{equation}
	Therefore, ${\rm Pr}\{{\rm error} \} = \sum_{\bm u^{K} \in \mathbb{F}_2^{K} } 2^{-K} {\rm Pr}\{{\rm error}|\bm u^K\} \leq  \epsilon$ and hence we have the first error probability
	\begin{equation}
		{\rm Pr}\{\bm u^{(0)}\neq \hat{\bm u}^{(0)}\}\leq \epsilon.
	\end{equation}
\par This completes the proof of Theorem~\ref{theorem_for_conv}.
\par\textbf{Remarks:} We shall see from the proof of Theorem~\ref{theorem_for_conv} that the first  error event probability of Conv-RaS codes goes to $0$ exponentially as $m\rightarrow\infty$. For practical applications, termination is needed, which can be achieved by padding $m$ sub-blocks of zeros at the tail of every, say, $m^3$, sub-blocks of data.  With this setup, the rate is reduced from $k/(n-k)$ to $m^2k/[(n-k)(1+m^2)]$ but the rate loss can be neglected for sufficiently large $m$.  In this case, the whole frame~(the $m^3$ sub-blocks) error probability is upper bounded by $m^3\cdot{\rm Pr}\{\bm u^{(0)}\neq \hat{\bm u}^{(0)}\}$~\cite{jacobs1965principles}, which can be made arbitrarily small. Also, our proof suggests that the totally random time-varying convolutional codes can achieve the capacity, which has been proved in~\cite{shulman2000improved}. Actually, if the elements of $\mathbf{G}$ are uniformly at random generated, a stronger conclusion can be made that the time-invariant random convolutional codes can achieve the capacity in terms of the first error event probability~\cite{Wang2023}.
\section{Simulation Results}\label{sec5}

 \par The time-invariant Conv-RaS codes are considered in the simulations. To describe the encoding more compactly and illustrate the flexibility of the construction, we define two transformations. One is the interleaving and repetition transformation $\mathbf{P}$. For a sub-block $\bm u^{(i)}$, we have $\bm u^{(i)}\mathbf{P}=[\bm u^{(i)}\mathbf{P}_0,\bm u^{(i)}\mathbf{P}_1 ,\cdots,\bm u^{(i)}\mathbf{P}_m]$, where $\mathbf{P}_i$ is a random permutation matrix. The other is the delay transformation $\mathbf{D}$. This transformation shifts one bit to the right and pads $0$ to the empty entry. Let $\bm u^{(i)}$, $0\leq i \leq M$, be the data for transmission. The codeword is then given by 
\begin{equation}
	\bm{x}=\sum_{{i}=0}^{{M}}\bm{{u}}^{({i})}\mathbf{P}\mathbf{D}^{i(n-k)}.
\end{equation}
where each sub-block $\bm u^{(i)}$ are interleaved/repeated $m+1$ times, shifted to right by $i(n-k)$ digits and then XORed for transmission. In real implementation, the encoding can be done sub-block by sub-block, which accepts $\bm u^{(i)}$ of length $k$ as input and delivers $\bm x^{(i)}$ of length $n-k$ as output. The decoding can be implemented by iterative sliding window decoding algorithm over a normal graph~\cite{Ma2015BMST,Cai2020SCLDGM}.
\par Notice that the the code is non-systematic and only $\bm x^{(i)}$ is transmitted. The code rate is defined as $k/(n-k)$. The Conv-RaS codes with rates less than one are typically used as channel codes for error correction, while the Conv-RaS codes with rates greater than one are typically used as source codes for data compression. To transmit biased sources over noisy channels, the Conv-RaS codes are used as joint source-channel codes with the code rate being adjusted according to the sparsity of the
source and the quality of the channel. To construct a Conv-RaS code, we need $m + 1$ random permutation matrices, where $m$ can be determined according to the target BER with the simple GA lower
bounds~\cite{Ma2015BMST,Cai2020SCLDGM}, which does not require complicated optimization.

\par In the following examples showing the flexibility and universality of Conv-RaS codes, we use the conventional rate $R = (n-k)/k$ for source coding and the transmission rate $R = k/(n-k)$ for channel coding and JSCC. 
\par
\begin{figure}[t]
	\centering
	\includegraphics[width=0.55\textwidth]{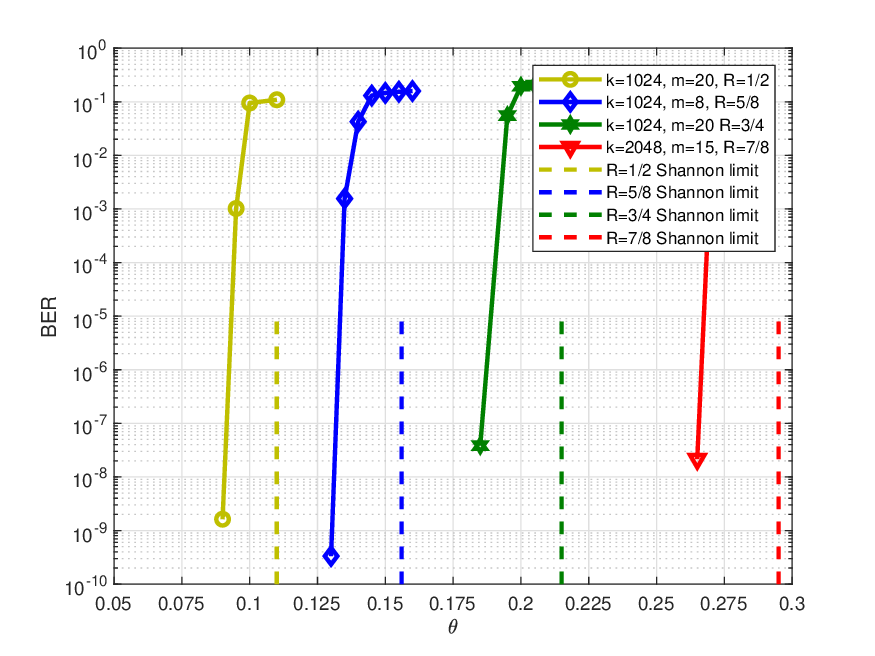}
	\caption{ The BER performance of Conv-RaS codes for source coding. Here $k \in \{1024,2048\}$, $n-k \in \{512,640, 768, 1792\}$ and $m\in\{8, 15,20\}$. The rate is defined as $R = (n-k)/k$ and the corresponding Shannon limit is defined as the maximum $\theta \leq  1/2$ such that $ H(\theta)\leq R$.  }
	\label{source_coding_BER}
\end{figure}

 \textbf{Example}~2~(Source coding only): In this example, we show the simulation results of Conv-RaS codes for source coding with Bernoulli source $\theta\triangleq{\rm Pr}\{U=1\}$. With fixed $k\in \{1024,2048\}$, the BER performance with different values of $m$ and $n-k$ and hence different  rates $R=(n-k)/k$ is shown in Fig.~\ref{source_coding_BER}. The  rates for Conv-RaS codes are $R\in\{ 1/2, 5/8, 3/4,7/8\}$ and the corresponding values of $\theta$ with $H(\theta)=R$ are also plotted. From the simulation results, we see that, to achieve the  BER of $10^{-5}$, the Conv-RaS codes can be close to the limits. 
\par
\begin{figure}[t]
	\centering
	\subfigure[The BER performance]{\includegraphics[width=0.5\textwidth]{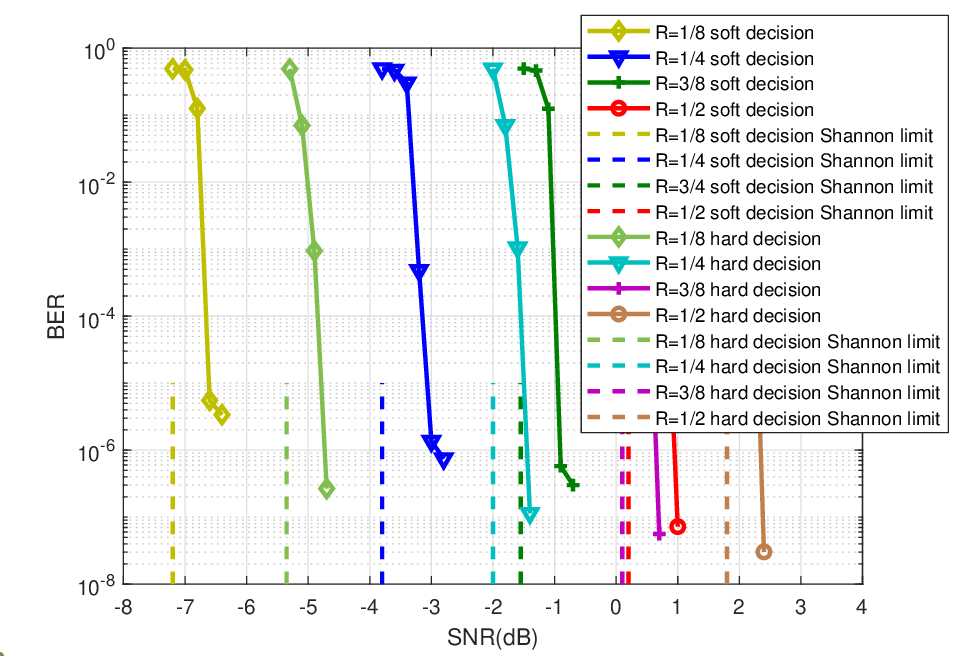}\label{channel_coding_BER}}
	\hspace{-0.2cm}
	\subfigure[The Conv-RaS codes to achieve the BER of~$10^{-5}$ ]{\includegraphics[width=0.45\textwidth]{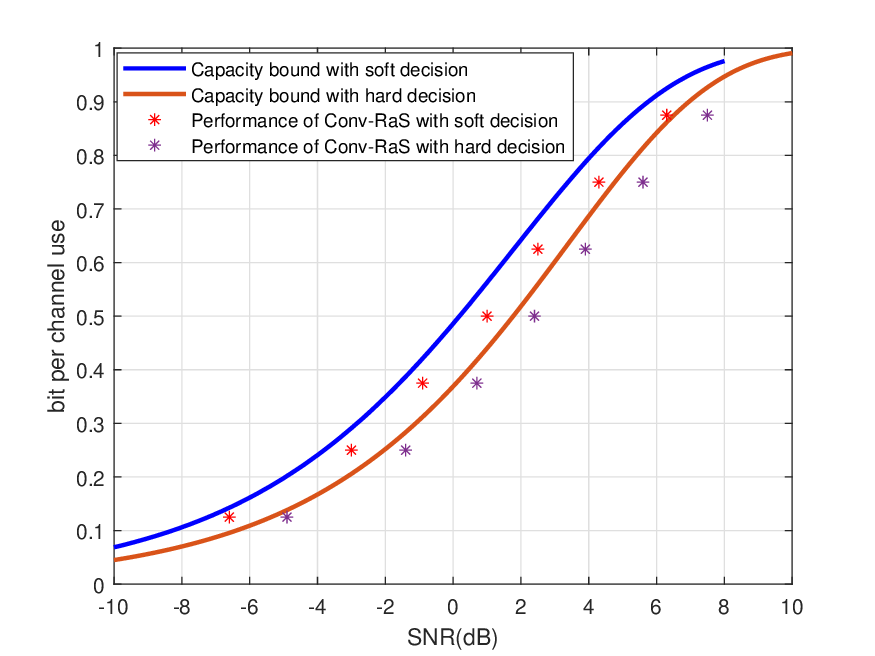}\label{channel_coding_limit}}
	\caption{The simulation results for channel coding with Conv-RaS codes. }
\end{figure}

 \textbf{Example}~3~(Channel coding only): In this example, we show the simulation results of Conv-RaS codes for channel coding with BPSK signalling over AWGN channels. The simulations for Conv-RaS codes are conducted with soft decision decoding and hard decision decoding~(corresponding to BSC). With fixed $k=1024$ and $m=10$, the BER performance with different values of $n-k$ and hence different code rates $R=k/(n-k)$ is shown in Fig.~\ref{channel_coding_BER}. The code rates for Conv-RaS codes are $R\in\{1/8$, $1/4$, $3/8, 1/2\}$ for both soft decision and hard decision. The corresponding Shannon limits for the code rates are also plotted. From the simulation results in Fig.~\ref{channel_coding_BER}, we see that the Conv-RaS can achieve the  BER of $10^{-5}$ within one dB from the Shannon limit. We also plot the required SNR to achieve a BER of $10^{-5}$ for the Conv-RaS codes with  $R\in \{1/8, 1/4, 3/8, 1/2, 5/8, 3/4, 7/8\}$ in Fig.~\ref{channel_coding_limit}, from which we see that the Conv-RaS codes can achieve the BER of $10^{-5}$   within one dB away from the Shannon limits for all code rates.
\begin{figure}[t]
	\centering
	\includegraphics[width=0.55\textwidth]{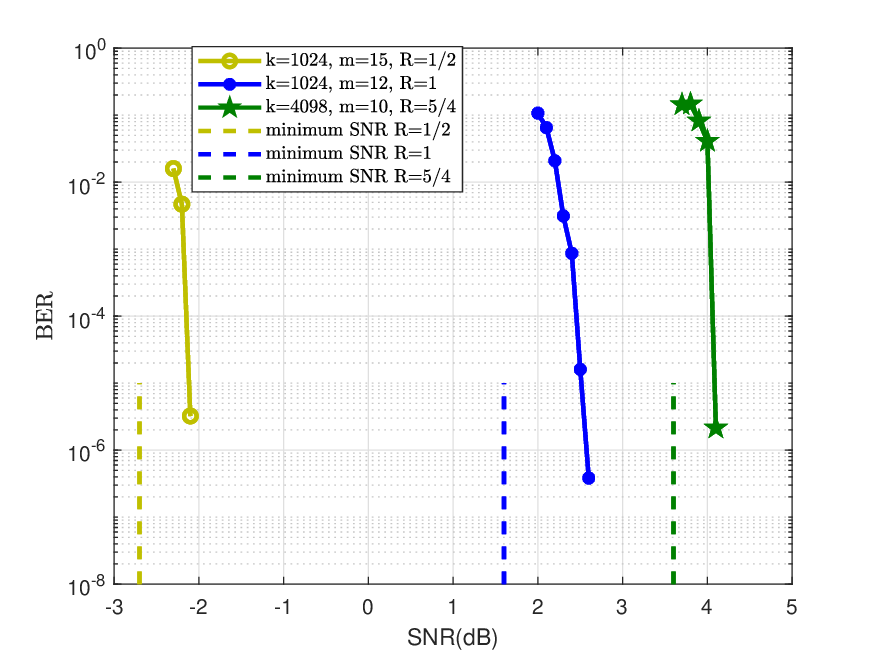}
	\caption{ The BER performance of $R\in\{1/2,1,5/4\}$,  $k\in \{1024,2048\}$, and fixed $\theta=0.15$. The corresponding limits are also plotted. Notice that the limit is the minimum SNR $= 1/(\sigma^2)$ satisfying $C$(SNR) $ > RH(\theta)$, where $C$(SNR) indicates that the capacity $C$ is  a function of SNR. The corresponding $E_b/N_0$ is given by  $E_b/N_0= {\rm SNR}/(2RH(\theta))$ and the required $E_b/N_0$ are $-0.55$~dB, $0.74$~dB and $1.78$~dB  for $R=1/2$, $1$ and $5/4$, respectively.}
	\label{JSCC_BER}
\end{figure}
 \par \textbf{Example}~4~(JSCC): In this example, we consider the Conv-RaS codes for JSCC with Bernoulli source of $\theta=0.15$ and BPSK signalling over AWGN channels. For the Conv-RaS codes, we set $k\in\{1024,2048\}$ and change the rates and values of $m$. The simulation results are shown in Fig.~\ref{JSCC_BER} and the values of code rates and $m$ are specified in the legends. From Corollary~\ref{corollary_JSCC}, we have $H(\theta)<C/R$, where $R=k/(n-k)$ is the code rate. Thus, we can derive the minimum required SNR for a given  $\theta$ and code rate $R$, which is also plotted in Fig.~\ref{JSCC_BER}.  From Fig.~\ref{JSCC_BER}, we see that the Conv-RaS codes can achieve the BER of $10^{-5}$ within one dB from the minimum required SNR. We also see that the Conv-RaS codes are flexible to achieve various code rates.
\par
\begin{figure}[t]
	\centering
	\includegraphics[width=0.55\textwidth]{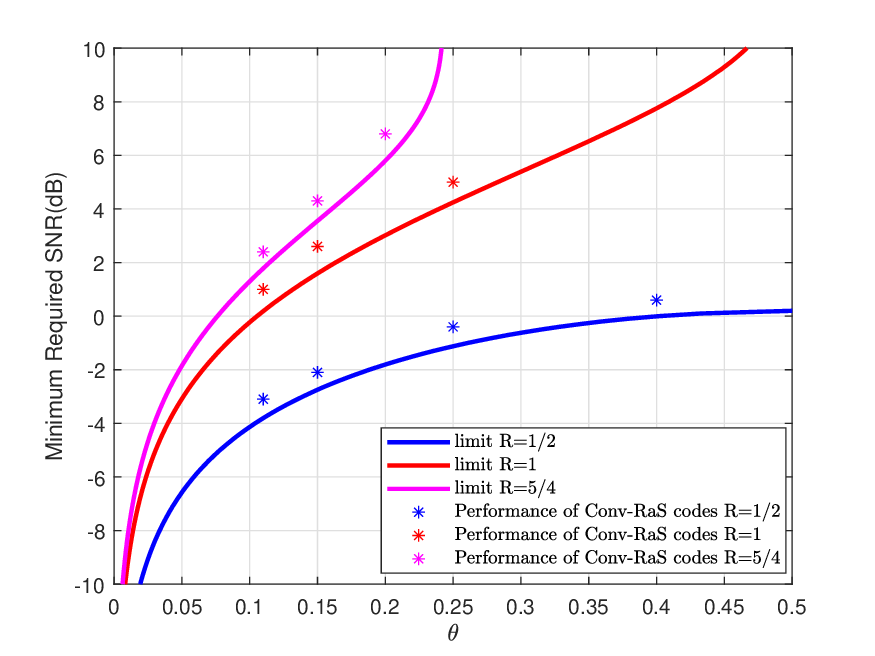}
	\caption{ The minimum required SNR by JSCC for different sources~($\theta$) and different code rates $R\in \{1/2, 1, 5/4\}$, where the theoretical limit is calculated by the condition that $C$(SNR) $ > RH(\theta)$ and the performance of  Conv-RaS codes is simulated at BER $= 10^{-5}$.	
  }
	\label{JSCC_limit}
\end{figure}

 \textbf{Example}~5: In this example, we show the performance of Conv-RaS codes with different $\theta$ and  $R\in\{1/2, 1, 5/4\}$. With BPSK signalling over AWGN channels, the performance for Conv-RaS codes to achieve BER of $10^{-5}$ are compared with the minimum required SNR for various code rates and $\theta$ in Fig.~\ref{JSCC_limit}. From Fig.~\ref{JSCC_limit}, we have the following observations, as expected.
\begin{itemize}
\item For a given code rate, the denser the source is, the higher the SNR is required.
\item For a given source~($\theta$), the higher the code rate is, the higher the SNR is required.
\item For a given channel~(SNR), the denser the source, the lower the code rate is required.
\item The Conv-RaS codes perform well and can achieve BER of $10^{-5}$ within one dB from the minimum required SNR.  
\end{itemize}

\section{Conclusions}
\label{sec6}
In this paper, we have proved that the  RaS code ensemble is capacity-achieving in terms of FER with the newly proposed framework. The coding theorem can be easily generalized to source coding and JSCC. To improve the performance,  we consider the Conv-RaS code ensemble with better performance under iterative decoding than their block counterpart and prove the Conv-RaS code ensemble is also capacity-achieving with the framework. Numerical results show that the  Conv-RaS codes can perform well in source coding, channel coding and JSCC and have flexibility in construction.
\section*{Acknowledgment}
The authors would like to thank Dr. Suihua Cai from Sun Yat-sen University  
for his helpful discussions.\\
\section*{Appendix}
\subsection{Proof of Lemma~\ref{lemma_for_rho}}
 Noticing that the parity-check vector of $\bm u^{(i)}\mathbf{G}_{i,j}$ for the sub-block $\bm u^{(i)}$ is a Bernoulli sequence with success probability $w_i/(k+1)$, the success probability given in~\eqref{rho} can be derived by induction. 
	
	\par For $m=0$, we have $\rho(\bm w^1)=w_0/(k+1)$ for $\bm w^1=(w_0)$.
	\par For induction, suppose that for $m=t-1$, we have $\rho(\bm w^{t})=\frac{1-\prod\limits_{i=0}^{t-1}(1-\frac{2w_i}{k+1})}{2}$ for $\bm w^t=(w_0,w_1,\cdots,w_{t-1})$. Then we have, for $m=t$,
	\begin{equation}
		\label{rho_0}
		\rho(\bm w^{t+1})=\rho(\bm w^{t})\left(1-\frac{w_t}{k+1}\right)+(1-\rho(\bm w^{t}))\frac{w_t}{k+1},
	\end{equation}
	\begin{equation}
		\label{rho_1}
		1-\rho(\bm w^{t+1})=\rho(\bm w^{t})\frac{w_t}{k+1}+(1-\rho(\bm w^{t}))\left(1-\frac{w_t}{k+1}\right),
	\end{equation} 
	where $\bm w^{t+1}=(w_0,w_1,\cdots,w_t)$.
	\par By subtracting~\eqref{rho_0} from~\eqref{rho_1}, we have 
	\begin{equation}
		1-2\rho(\bm w^{t+1})=(1-2\rho(\bm w^{t}))\left(1-\frac{2w_t}{k+1}\right).	
	\end{equation} 
	\par By induction hypothesis, we have 
	$$\rho(\bm w^{t+1})=\frac{1-\prod\limits_{i=0}^{t}(1-\frac{2w_i}{k+1})}{2}\text{,}$$
	which proves the first statement of this lemma. To complete the proof, we assume without loss of generality that the first $T$ sub-blocks in $\bm u$ are non-zero sub-blocks. For the zero sub-blocks, we have  $1-2w_i/(k+1)=1$ since $w_i=0$. As a result, we have
	\begin{equation}
		\label{rho_T_1}
		\rho(\bm w^{m+1})=\rho(\bm w^{T})=\frac{1-\prod\limits_{i=0}^{T-1}(1-\frac{2w_i}{k+1})}{2}.
	\end{equation}
	
	For the $i$-th non-zero sub-block in $\bm u$ with Hamming weight $w_i$, we define $\theta_i=1/2-w_i/(k+1)$. From~\eqref{rho_T_1}, we have 
	\begin{equation}
		\left|\rho(\bm w^{T}) -\frac{1}{2}\right|=2^{T-1}\prod\limits_{i=0}^{T-1}|\theta_i|.
	\end{equation}
	With $1\leq w_i\leq k$, from the definition of $\theta_i$, we have 
	\begin{equation}
		|\theta_i|=\left|\frac{w_i}{k+1}-\frac{1}{2}\right|\leq  \frac{k-1}{2(k+1)}.
	\end{equation}
	Thus, we have 
	\begin{equation}
		\begin{aligned}
			\left|\rho(\bm w^{T}) -\frac{1}{2}\right|&=2^{T-1}\prod\limits_{i=0}^{T-1}|\theta_i|\\
			&\leq 2^{T-1}\left[\frac{k-1}{2(k+1)}\right]^{T}=\frac{1}{2}\left(\frac{k-1}{k+1}\right)^{T}.
		\end{aligned}
	\end{equation}
	Therefore,
	\begin{equation}\label{rho_bound}
		\frac{1}{2}-\frac{1}{2}\left(\frac{k-1}{k+1}\right)^{T}\leq \rho(\bm w^{T})\leq \frac{1}{2}+\frac{1}{2}\left(\frac{k-1}{k+1}\right)^{T}.
	\end{equation}
	Thus, for $T\geq 1$, we have 
	\begin{equation}\label{lower_bound_for_rho}
		0<\frac{1}{2}-\frac{k-1}{2(k+1)}\leq \rho(\bm w^T)=\rho(\bm w^{m+1}).
	\end{equation}
	Hence, from Lemma~\ref{partial_mutual_information}, we have
	\begin{equation}
		I_0(\rho(\bm w^{m+1}))=I_0(\rho(\bm w^{T}))\geq I_0\left(\frac{1}{2}-\frac{k-1}{2(k+1)}\right)>0.
	\end{equation} 
	From~\eqref{rho_bound}, for any $\bm u\in \mathbb{F}_{2}^{K}$ with $t\geq T$ non-zero sub-blocks, we have
	\begin{equation}
		\begin{aligned}
			P(\bm x|\bm u)&\triangleq {\rm Pr}\{\bm X=\bm x|U=\bm u \}\leq\left[\frac{1}{2}+\frac{1}{2}\left(\frac{k-1}{k+1}\right)^{T}\right]^{N-K}\text{.}
		\end{aligned}
	\end{equation}

\bibliographystyle{IEEEtran}
\bibliography{bibliofile} 

\end{document}